\documentclass[paper]{JHEP3}

\usepackage{amsmath,amssymb}
\usepackage{graphicx}

\newcommand{\be}{\begin{equation}}
\newcommand{\ee}{\end{equation}}
\def\tr{{\rm tr}\,}
\def\Tr{{\rm Tr}\,}
\def\cN{{\cal N}}
\def\cD{{\cal D}}

\def\bea{\begin{eqnarray}}
\def\eea{\end{eqnarray}}
\def\nn{\nonumber}
\def\nnabla{\nabla\hspace{-8pt}\nabla}

\preprint{ITP--UH--14/09\\
JINR-E2-2009-141}

\title{Quantum  $\cN{=}3$, $d{=}3$ Chern-Simons matter theories\\
in harmonic superspace}

\author{I.L. Buchbinder\\
    Department of Theoretical Physics, Tomsk State
Pedagogical University, 634041 Tomsk, Russia,
    E-mail: \email{joseph@tspu.edu.ru}}

\author{E.A. Ivanov\\
Bogoliubov Laboratory of Theoretical Physics, JINR, 141980 Dubna,
Russia \\
E-mail: \email{eivanov@theor.jinr.ru}}

\author{O. Lechtenfeld\\
Institut f\" ur Theoretische Physik, Leibniz Universit\" at
Hannover, 30167 Hannover, Germany, E-mail:
\email{lechtenf@itp.uni-hannover.de}}

\author{N.G. Pletnev\\
Department of Theoretical Physics, Institute of Mathematics,
Novosibirsk, 630090, Russia, E-mail: \email{pletnev@math.nsc.ru}}

\author{I.B. Samsonov\\
Laboratory of Mathematical Physics, Tomsk Polytechnic University,
634050 Tomsk, Russia, E-mail:
\email{samsonov@mph.phtd.tpu.edu.ru}}

\author{B.M. Zupnik\\
Bogoliubov Laboratory of Theoretical Physics, JINR, 141980 Dubna, Russia \\
E-mail: \email{ zupnik@theor.jinr.ru}}

\abstract{We develop the background field method for studying
classical and quantum aspects of $\cN{=}3$, $d{=}3$ Chern-Simons
and matter theories in $\cN{=}3$ harmonic superspace. As one of
the immediate consequences, we prove a nonrenormalization theorem
implying the ultra-violet finiteness of the corresponding
supergraph perturbation theory. We also derive the general
hypermultiplet and gauge superfield propagators in a Chern-Simons
background. The leading supergraphs with two and four external
lines are evaluated. In contrast to the non-supersymmetric theory,
the leading quantum correction to the massive charged
hypermultiplet proves to be the super Yang-Mills action rather
than the Chern-Simons one. The hypermultiplet mass is induced by a
constant triplet of central charges in the $\cN{=}3$, $d{=}3$
Poincar\'e superalgebra.}


\begin{document}
\section{Introduction}
Three-dimensional extended supersymmetric gauge theories attract much attention
due to their remarkable relationships with string/M theory. In this paper we develop
a generic procedure for constructing quantum effective actions of $\cN{=}3$,
$d{=}3$ supergauge theories in terms of unconstrained harmonic
superfields.

The harmonic superspace approach~\cite{GIKOS,GIOS1,GIOS2,book} is
a powerful tool for studying field theories with extended
supersymmetry in diverse dimensions. In particular, the $\cN{=}3$,
$d{=}3$ harmonic superspace was introduced in \cite{5}.
Recently~\cite{BILPSZ}, we applied this approach to the
three-dimensional $\cN{=}3$ supersymmetric Chern-Simons and matter
models which are building blocks of the $\cN{=}6$ and $\cN{=}8$
supersymmetric Aharony-Bergman-Jafferis-Maldacena
(ABJM)~\cite{ABJM} and Bagger-Lambert-Gustavsson (BLG)~\cite{BLG}
theories. The manifestly $\cN{=}3$ supersymmetric off-shell
formulation of these theories was constructed for the first time.
The ABJM and BLG models are currently of great interest, because
they describe the world-volume dynamics of M2 branes in
superstring theory and so open a way for studying the
AdS$_4$/CFT$_3$ correspondence.

Superspace formulations are most advantageous for studying the quantum aspects
of supersymmetric field theories  because they make manifest one or another amount of the underlying supersymmetries.
Based on this general feature, it is natural to expect that the $\cN{=}3$, $d{=}3$
harmonic superspace approach may prove very fruitful for
quantum computations in the $\cN{=}6$ or $\cN{=}8$
superconformal models including the ABJM and BLG ones.
However, there is very limited experience in quantizing
$\cN{=}3$, $d{=}3$ Chern-Simons or matter models directly in
harmonic superspace \cite{5} (as opposed to e.g.\ $\cN{=}2$, $d{=}4$
supersymmetric theories \cite{GIOS1,GIOS2,book}).
The aim of the present paper is to partially fill this gap by working
out the basic steps of the appropriate quantization procedure.

We develop the background field method for
a general $\cN{=}3$ Chern-Simons matter theory
and use it to prove a nonrenormalization theorem which
guarantees the quantum finiteness of such a theory.
We derive the superfield propagators in these models and use them for
the calculation of leading supergraphs with two and four external
gauge and matter legs. These diagrams bring to light some interesting
facts about the quantum theory. First,
the leading quantum correction in the massive charged hypermultiplet
generates the $\cN{=}3$ super Yang-Mills action rather than
the Chern-Simons term. This result is rather unexpected in comparison with
$\cN{=}0$ three-dimensional electrodynamics in which a single massive
fermion generates the Chern-Simons action as a leading contribution
to the effective action~\cite{Redlich}.
Second, the four hypermultiplet one-loop diagram produces a quartic
hypermultiplet self-interaction.
Like in the four-dimensional case~\cite{IKZ}, such a contribution
is possible only in a model of massive charged hypermultiplets, with
the hypermultiplet mass being induced by the central charge of
the $\cN{=}3, d{=}3$ Poincar\'e superalgebra.

The paper is organized as follows. In Section~2 we review the
formulation of the $\cN{=}3$ hypermultiplet and Chern-Simons
models in $\cN{=}3$ harmonic superspace. In Section~3 we develop
the background field method for a general $\cN{=}3$ Chern-Simons
matter theory, prove the nonrenormalization theorem and discuss
the general structure of the quantum effective action. In
Section~4 we consider some one-loop quantum computations in the
case of vanishing background field. In Section~5 we study the
realization of $\cN{=}3$ supersymmetry with a central charge. This
leads us to the massive hypermultiplet model, whose quantum
aspects we consider as well. The final Section~6 contains a
discussion of our results as well as prospects of their further
applications to three-dimensional models with extended
supersymmetry. In the Appendix we collect technical details of the
$\cN{=}3$, $d{=}3$ harmonic superspace approach.

\section{Field models in $\cN{=}3$, $d{=}3$ harmonic superspace}
The basic aspects of the $\cN{=}3$, $d{=}3$ harmonic superspace were
worked out for the first time  in \cite{5}. In this paper we follow the notations
used in our recent paper \cite{BILPSZ}. They are collected in the
Appendix.

\subsection{Gauge theory in standard $\cN{=}3$ superspace}
To begin with, we consider the gauge theory in standard
$\cN{=}3$, $d{=}3$ superspace with coordinates
$(x^{\alpha\beta},\theta_\alpha^{ij})$ and covariant spinor derivatives
$D_\alpha^{ij}$ given by (\ref{Q}). Following the standard geometric
approach to gauge theories in superspace, we start by defining
the superfield connections for the space-time and spinor
derivatives,
\be
\nabla^{ij}_\alpha=D^{ij}_\alpha+V^{ij}_\alpha\,,\qquad
\nabla_{\alpha\beta}=\partial_{\alpha\beta}+V_{\alpha\beta}\,.
\ee
The main superfield constraint for these superfield connections
is given by \cite{5}
\be
\{\nabla^{ij}_\alpha,\nabla^{kl}_\beta \}=i\nabla_{\alpha\beta}
(\varepsilon^{ik}\varepsilon^{jl}+\varepsilon^{il}\varepsilon^{jk})
-\frac12\varepsilon_{\alpha\beta}(\varepsilon^{ik}W^{jl}
+\varepsilon^{il}W^{jk}+\varepsilon^{jl}W^{ik}+\varepsilon^{jk}W^{il})\,,
\label{main}
\ee
where $W^{ij}=W^{(ij)}$ is a superfield strength for these gauge
connections. Using the Bianchi identities one can check that the
commutators of other covariant derivatives do not involve new
tensors except $W^{ij}$ and its derivatives,
\bea
[\nabla_{\alpha\beta}, \nabla^{ij}_\gamma]
&=&\varepsilon_{\alpha\gamma}F^{ij}_\beta
+\varepsilon_{\beta\gamma}F^{ij}_\alpha\,,
\\
{}[\nabla_{\alpha\beta},\nabla_{\gamma\delta}]
&=&\varepsilon_{\alpha\gamma}F_{(\beta\delta)}
+\varepsilon_{\beta\gamma}F_{(\alpha\delta)}
+\varepsilon_{\beta\delta}F_{(\alpha\gamma)}
+\varepsilon_{\alpha\delta}F_{(\beta\gamma)}\,,
\eea
where
\be
F_\alpha^{ij}=\frac i4(\nabla_\alpha^{ik}W^j_k+\nabla_\alpha^{jk}W^i_k)\,,
\qquad
F_{(\alpha\beta)}
=\frac1{24}(\nabla_{\alpha\, ij}\nabla_{\beta}^{ik}W_k^j
+\nabla_{\beta\, ij}\nabla_\alpha^{ik}W_k^j)\,.
\label{F}
\ee
Moreover, the Bianchi identities lead to the following off-shell
constraint for $W^{ij}$,
\be
\nabla^{(ij}_\alpha W^{kl)}=0\,.
\label{constr}
\ee
In the next subsection we will show how this constraint is
resolved within the harmonic superspace approach.

\subsection{Gauge theory in $\cN{=}3$ harmonic superspace}

The $\cN{=}3$ harmonic superspace is parametrized by the following
coordinates \footnote{Note that in \cite{5} the $\cN{=}3$, $d{=}3$ harmonic
superspace with ${\rm O}(3)/{\rm O}(2)$ harmonics was introduced.}
\be
z=\{x^{\alpha\beta},\theta^{++}_\alpha,\theta^{--}_\alpha,\theta^0_\alpha,
u^\pm_i \}\,, \label{Central}
\ee
where $\theta^{\pm\pm}_\alpha=\theta^{ij}_\alpha
u^\pm_i u^\pm_j$, $\theta^0_\alpha=\theta^{ij}_\alpha u^+_i u^-_j$
and $u^\pm_i$ are the ${\rm SU}(2)/{\rm U}(1)$ harmonic coordinates
subjected to the constraints $u^{+i}u^-_i=1$, $u^{+i}u^+_i=0$,
$u^{-i}u^-_i=0$. The harmonic projections of the
covariant spinor derivatives $\nabla^{ij}_\alpha$ and the
superfield strengths $W^{ij}$ are defined as follows
\bea
&&\nabla^{++}_\alpha=u^+_i
u^+_j \nabla^{ij}_\alpha\,,\quad
\nabla^{--}_\alpha=u^-_i u^-_j
\nabla^{ij}_\alpha\,,\quad
\nabla^0_\alpha=u^+_i u^-_j \nabla^{ij}_\alpha\,,\nn\\
&&W^{++}=u^+_i u^+_j W^{ij}\,,\quad
W^{--}=u^-_i u^-_j W^{ij}\,,\quad
W^0=u^+_i u^-_j W^{ij}\,.
\eea
There are obvious relations between the harmonic superfield strengths,
\be
({\rm a}) \;\;\partial^{++}W^{++}=0\,,\quad
({\rm b}) \;\; W^0=\frac12 \partial^{--}W^{++}\,,\quad
W^{--}=\partial^{--} W^0=\frac12(\partial^{--})^2W^{++}\,,
\label{W0}
\ee
where harmonic derivatives in the central basis $\partial^{\pm\pm}$ are defined in Appendix.
In terms of the above harmonic projections, the anticommutation relations (\ref{main}) can be rewritten as
\bea
\{\nabla^{++}_\alpha,\nabla^{--}_\beta
\}&=&2i\nabla_{\alpha\beta}+2\varepsilon_{\alpha\beta}W^0\,,\qquad
\{\nabla^0_\alpha,\nabla^0_\beta \}=-i\nabla_{\alpha\beta}\,,
\nn\\
\{\nabla_\alpha^{++},\nabla^0_\beta
\}&=&\varepsilon_{\alpha\beta}W^{++}\,,\qquad
\{\nabla_\alpha^{--},\nabla^0_\beta
\}=-\varepsilon_{\alpha\beta}W^{--}\,,
\label{2.11}
\eea
while the harmonic projections of the constraint (\ref{constr})
are given by
\bea
&&\nabla_\alpha^{++}W^{++}=0\,,\label{Basic} \\
&& \nabla_\alpha^{--}W^{--}=0\,,\quad
\nabla_\alpha^{--}W^{++}+4\nabla^0_\alpha W^0 +\nabla^{++}_\alpha W^{--}=0\,,\nn\\
&&\nabla^0_\alpha W^{++}+\nabla^{++}_\alpha W^0=0\,,\qquad
\nabla^0_\alpha W^{--}+\nabla^{--}_\alpha W^0=0\,.
\label{constraint1}
\eea
The relations (\ref{Basic}) and (\ref{constraint1}) are none other than the Bianchi
identities for the superfield strengths $W^{++}$, $W^{--}$, $W^0$. It is important to realize
that the whole set of the constraints (\ref{constraint1}) can be produced from the relation (\ref{Basic}) by
the successive action of the harmonic derivative $\partial^{--}$. Thus eq.\ (\ref{Basic}) is the basic constraint.
As will be clear soon, it is nothing else as the Grassmann analyticity condition, and it can be
solved by passing to the analytic basis in ${\cal N}{=}3, d{=}3$ harmonic superspace and to an analytic gauge frame.

An important feature of the $\cN{=}3$, $d{=}3$ harmonic superspace  is the
existence of an analytic subspace in it. This subspace is closed under the $\cN{=}3$ supersymmetry
and is parametrized by the following coordinates
\be
\zeta_A=(x^{\alpha\beta}_A,
\theta^{++}_\alpha, \theta^{0}_\alpha, u^\pm_i)\,,
\ee
where
\be
x^{\alpha\beta}_A=(\gamma_m)^{\alpha\beta}x^m_A=x^{\alpha\beta}
+i(\theta^{\alpha++}\theta^{\beta--}+\theta^{\beta++}\theta^{\alpha--})\,.
\ee
The analytic basis of $\cN{=}3$, $d{=}3$ harmonic superspace (as opposed to the
original, central basis $(x^{\alpha\beta},\theta_\alpha^{ij})$)
is defined as the coordinate set
\be
z_A = \{\zeta_A, \theta^{--}_\alpha \}\,. \label{Analytic}
\ee
In the analytic basis the Grassmann derivative $D^{++}_\alpha$ becomes
short, $D^{++}_\alpha=\tfrac\partial{\partial\theta^{--\alpha}}$.
Other Grassmann and harmonic derivatives in this basis are given by expressions (\ref{A16}),( \ref{A17}).
The existence of the analytic subspace is crucial for constructing superfield actions, as it
allows one to define the analytic (short) superfields, which are
independent of the coordinate $\theta^{--\alpha}$,
\be
D^{++}_\alpha\Phi_A=0 \quad \Rightarrow \quad \Phi_A = \Phi_A(\zeta_A)\,.
\label{PhiA}
\ee

As soon as the harmonic variables $u^\pm_i$ appear on equal
footing with the other superspace coordinates, there is a set of the
harmonic derivatives $\partial^{++}$, $\partial^{--}$,
$\partial^0$ given by (\ref{A12}). Clearly, these
derivatives do not receive any gauge connections in the original gauge frame (``$\tau$ frame''), since the gauge
transformations in it are associated with the harmonic-independent,
gauge algebra valued superfield parameter $\tau=\tau(z)$,
e.g.,
\be
W^{ij}\longrightarrow e^{\tau}W^{ij} e^{-\tau}\,,\qquad
\partial^{++}\tau=\partial^{--}\tau=0\,.
\ee
However, in order to be able to deal with the manifestly analytic superfields (\ref{PhiA})
in non-trivial representations of the gauge group, one should define another gauge frame (``$\lambda$ frame'')
in which the gauge group is represented by the analytic superfield transformations,
\be
\Phi_A\longrightarrow \Phi'_A= e^\lambda \Phi_A\,,\qquad
D^{++}_\alpha\lambda=0\,.
\ee
Having two different representations of the same gauge group, one with
the harmonic-independent gauge parameter $\tau$ and another
with the analytic gauge parameter $\lambda$, one can define the invertible ``bridge''
$e^\Omega$, $\Omega=\Omega(z,u)$, which transforms as
\be
e^{\Omega'}=e^\lambda e^\Omega e^{-\tau}\,,
\ee
and thus relates the $\tau$ and $\lambda$ frames \cite{GIKOS}:
\be
\Phi_{A(\tau)}=e^{-\Omega}\Phi_A\,,\qquad
\Phi_{A(\tau)}\longrightarrow \Phi'_{A(\tau)}= e^\tau \Phi_{A(\tau)}\,.
\ee
Respectively, the Grassmann and harmonic
gauge covariant derivatives in the $\tau$ and $\lambda$ frames are related as
\bea
\nabla^{++}_{(\lambda)\alpha} &=& e^{-\Omega}\nabla^{++}_\alpha
e^{\Omega}=D^{++}_\alpha\,,\quad
\nabla^{--}_{(\lambda)\alpha} =
 e^{-\Omega}\nabla^{--}_\alpha e^{\Omega}\,,\quad
\nabla^{0}_{(\lambda)\alpha} = e^{-\Omega}\nabla^{0}_\alpha
e^{\Omega}\,,\label{bridge0}
\\
\nabla^{\pm\pm}_{(\lambda)}&=&e^{-\Omega}{\cal D}^{\pm\pm}e^{\Omega}
= {\cal D}^{\pm\pm}+V^{\pm\pm}\,,\qquad
 V^{\pm\pm}=e^{-\Omega}({\cal D}^{\pm\pm}e^{\Omega})\,,
\label{bridge}
\eea
where ${\cal D}^{\pm \pm}$ are the analytic-basis harmonic derivatives defined in (\ref{A16}).\footnote{We would equally choose
the central-basis form of the harmonic derivatives in (\ref{bridge0}), because there is no direct correlation between the superspace
bases and the gauge frames.}
Hereafter, we omit the subscript $(\lambda)$, assuming that we will
always make use of the $\lambda$ frame.

It is crucial that the derivative $\nabla^{++}_\alpha$ in the
$\lambda$ frame becomes short while the harmonic derivatives
acquire gauge connections. Owing to the commutation relation
$[D^{++}_\alpha,\nabla^{++}]=0$, the superfield $V^{++}$ is
analytic,
\be
D^{++}_\alpha V^{++}=0\,.
\ee
The algebra of harmonic derivatives $[\nabla^{++},\nabla^{--}]={\cal D}^0$
leads to the harmonic zero-curvature equation,
\be
{\cal D}^{++}V^{--}-{\cal D}^{--}V^{++}+[V^{++},V^{--}]=0\,,
\label{zero-curv}
\ee
which defines the gauge prepotential $V^{--}$ as a function of
$V^{++}$. An explicit solution of this equation can be represented
by the series \cite{BMZ87},
\be
V^{--}(z,u)=\sum_{n=1}^\infty(-1)^n\int du_1\ldots du_n
\frac{V^{++}(z,u_1)V^{++}(z,u_2)\ldots V^{++}(z,u_n)}{
(u^+u^+_1)(u^+_1u^+_2)\ldots(u^+_n u^+)}\,.
\label{V--}
\ee
It is important that not only the prepotential $V^{--}$,
but the gauge connections for the Grassmann derivatives, as well as the superfield
strengths in (\ref{2.11}), can be expressed through $V^{++}$. In
particular,
\bea
[\nabla^{--},D^{++}_\alpha]=2\nabla^0_\alpha
&\quad\Rightarrow\quad& V^0_\alpha=-\frac12 D^{++}_\alpha V^{--}\,,
\label{V-0}\\
\{D^{++}_\alpha,\nabla^0_\beta \}=\varepsilon_{\alpha\beta}W^{++}
&\Rightarrow&W^{++}=-\frac14 D^{++\alpha}D^{++}_\alpha V^{--}\,,
\label{W++}
\eea
where $V^{--}$ is a function of $V^{++}$ given by (\ref{V--}).
The equations (\ref{W0}b), being rewritten in the $\lambda$ frame, read
\be
W^0=\frac12 \nabla^{--}W^{++}\,,\qquad
W^{--}=\nabla^{--} W^0=\frac12(\nabla^{--})^2W^{++}\,.
\label{W0_}
\ee

The equation (\ref{Basic}), when written in the $\lambda$
frame, just means the analyticity of the superfield strength
$W^{++}$,
\be
D^{++}_\alpha W^{++}=0
 \quad \Rightarrow \quad W^{++} = W^{++}(\zeta_A)\,.
\label{analW}
\ee
As a result, the superfield constraint (\ref{Basic}) for $W^{++}$ is
solved by using the $\lambda$ representation of the
gauge group and the analytic basis (\ref{Analytic}) in the harmonic superspace. However, the relation $\partial^{++}
W^{++}=0$, eq. (\ref{W0}a), which, in the $\tau$ frame, just states that $W^{++}$ is homogeneous of degree 2 in $u^+_i$,
becomes non-trivial in the $\lambda$ frame:
\be
\nabla^{++}W^{++}=0\,.
\ee
In particular, this constraint encodes the Bianchi identity for
the gauge field component of the gauge superfield strength.

\subsection{$\cN{=}3$ super Yang-Mills and Chern-Simons models}
As shown in the previous subsection, the $\cN{=}3$, $d{=}3$ gauge
theory is described by the superfield strengths $W^{++}$, $W^0$,
$W^{--}$ which can be expressed through the single analytic gauge
prepotential $V^{++}$. Since the superfield $W^{++}$ is analytic,
eq. (\ref{analW}), it can be used for constructing the actions directly in the
analytic subspace. In particular, the super Yang-Mills (SYM) and
Chern-Simons actions are given by \cite{5}
\bea
S_{SYM}&=&\frac1{g^2}\tr\int d\zeta^{(-4)}W^{++}W^{++}\,,
\label{SYM}\\
S_{CS}&=&\frac{ik}{4\pi}\tr\sum_{n=2}^\infty
\frac{(-1)^n}{n}
 \int d^3x d^6\theta du_1\ldots du_n
 \frac{V^{++}(z,u_1)\ldots V^{++}(z,u_n)}{(u^+_1u^+_2)\ldots (u^+_n
 u^+_1)}\,.
\label{CS}
\eea
Here $g$ is the Yang-Mills coupling constant with the mass
dimension $[g]=1/2$ while $k$ is the (integer) Chern-Simons level.
The rules of integration over the analytic and full $\cN{=}3$
superspaces are given in the Appendix. Both SYM and Chern-Simons
actions are invariant under the following gauge
transformations
\be
V^{++}\longrightarrow {V^{++}}'=e^\lambda
\nabla^{++}e^{-\lambda}\,,
\label{gauge-tr-fin}
\ee
or, in the infinitesimal form,
\be
\delta_\lambda V^{++}=-\nabla^{++}\lambda=-{\cal D}^{++}\lambda
-[V^{++},\lambda]\,,
\label{gauge-tr}
\ee
where $\lambda$ is an analytic gauge parameter.

One can partly fix the gauge freedom by passing to the
Wess-Zumino gauge, in which
\bea
V^{++}_{WZ}&=&3(\theta^{++})^2u^-_k u^-_l \phi^{kl}(x_A)
+2\theta^{++\alpha}\theta^{0\beta}A_{\alpha\beta}(x_A)
+2(\theta^0)^2\theta^{++\alpha}\lambda_\alpha(x_A)\nn\\&&
+3(\theta^{++})^2\theta^{0\alpha}u^-_k
u^-_l\chi_\alpha^{kl}(x_A) +3i(\theta^{++})^2(\theta^0)^2u^-_ku^-_l
X^{kl}(x_A)\,.
\label{V++}
\eea
Such a form of the gauge prepotential is most suitable for
deriving the component structure of the SYM and Chern-Simons actions
\footnote{In the component field formulation such actions were
obtained in \cite{Kao}.}
\bea
S_{SYM}&=&\frac1{g^2}\tr\int d^3x\Big(
\phi^{kl}\square\phi_{kl}+\frac14 f^{\alpha\beta}f_{\alpha\beta}
- i\lambda^\alpha\partial_{\alpha\beta}\lambda^\beta
-\frac i2\chi^{kl\alpha}\partial_{\alpha\beta}\chi^\beta_{kl}
-X^{kl}X_{kl}
\nn\\&&+\mbox{interaction}
 \Big)\,,
\label{SYM-comp}
\\
S_{CS}&=&\frac{k}{4\pi}\tr\int d^3x\Big(
\phi^{kl} X_{kl}
-\frac{2i}3\phi^i_j[\phi^k_i,\phi^j_k]
+\frac i2\lambda^\alpha\lambda_\alpha
-\frac i4\chi^\alpha_{kl}\chi^{kl}_\alpha\nn\\&&
-\frac12A^{\alpha\beta}\partial^\gamma_{\alpha}A_{\beta\gamma}
-\frac{i}6
A^\alpha_\beta[A^\gamma_\alpha,A^\beta_\gamma] \Big)\,,
\label{CS-comp}
\eea
where $f_{\alpha\beta}=\partial_\alpha^\gamma A_{\beta\gamma}
+\partial_\beta^\gamma A_{\alpha\gamma}$.

Since the SYM model in three-dimensional space-time involves the
dimensionful coupling constant, it is not superconformal. In contrast,
the Chern-Simons theory has dimensionless coupling constant and therefore is
superconformal. In this paper we will be basically interested in
quantum aspects of superconformal models; so our main focus will
be on the $\cN{=}3$ Chern-Simons gauge theory, rather than on $\cN{=}3$ SYM. One more class
of $\cN{=}3$ superconformal theories we shall study is those of  matter hypermultiplets.

\subsection{$\cN{=}3$ hypermultiplets}
\label{sec2.4}
There are two basic types of the hypermultiplet in four
dimensions: the $q$ hypermultiplet and the $\omega$ hy\-per\-mul\-ti\-plet.
They describe the same physical degrees of freedom, though with different
assignments with respect to the R symmetry SU(2) group. Quite analogously,
both these types of hypermultiplets exist in three-dimensional space-time too.
In particular, the $q$-hypermultiplet consists of a ${\rm SU}(2)$
doublet of complex scalars $f^i$ and a doublet of complex spinors
$\psi^i_\alpha$ on shell. These fields appear in the component
expansions of the complex analytic superfield $q^+$ as
\be
q^+=u^+_i
f^i+(\theta^{++\alpha}
u^-_i-\theta^{0\alpha}u^+_i)\psi^i_\alpha
-2i(\theta^{++\alpha}\theta^{0\beta})\partial^A_{\alpha\beta}f^iu^-_i
+\mbox{aux. fields}.
\label{hyp-comp}
\ee
The free hypermultiplet action has the well known form,
\be
S_q= \int d\zeta^{(-4)} \bar q^+{\cal D}^{++}q^+\,, \quad \bar q^+ = \widetilde{q^+}\,, \;\;
\widetilde{\bar q^+} = -q^+\,,\label{qFree}
\ee
which yields the following free action for the physical components
upon eliminating an infinite tower of the auxiliary fields:
\be
S_{q,phys}
= -\int d^3x(\bar f_i \square f^i
+\frac i2\bar \psi^{\alpha}_i
\partial_{\alpha\beta}\psi^{i\beta})\,.
\ee

The $\omega$-hypermultiplet collects on shell a real scalar
$\varphi$, a triplet of real scalars $\varphi^{(ij)}$, a real
spinor $\psi_\alpha$ and a triplet of real spinors $\psi^{(ij)}_\alpha$
which appear in the component expansion of a real analytic
superfield $\omega$ as
\bea
\omega&=&\frac12\varphi+\frac1{\sqrt2}\varphi^{ij}u^+_iu^-_j
+\frac i{\sqrt2}\theta^{0\alpha}\psi_\alpha
+\frac{i\sqrt 3}2\theta^{++\alpha}\psi^{ij}_\alpha u^-_i u^-_j
-i\sqrt3\theta^{0\alpha}\psi^{ij}_\alpha u^+_i u^-_j
\nn\\&&
-i\sqrt2 \theta^{++\alpha}\theta^{0\beta}
\partial_{\alpha\beta}\varphi^{ij}u^-_i u^-_j
+\mbox{aux. fields}.
\eea
The free superfield action
\be
S_\omega=\int d\zeta^{(-4)} {\cal D}^{++}\omega {\cal
D}^{++}\omega\,,\qquad \tilde \omega=\omega\,,
\label{Om}
\ee
gives the standard kinetic terms for the physical component fields,
\be
S_{\omega,phys}=-\frac12\int d^3x(\varphi\square\varphi
+\varphi^{ij}\square\varphi_{ij}
+i\psi^\alpha\partial_{\alpha\beta}\psi^\beta
+i\psi^{ij\,\alpha}\partial_{\alpha\beta}\psi^\beta_{ij})\,.
\ee

The minimal gauge interaction of hypermultiplets can be implemented
by promoting the flat harmonic derivative ${\cal D}^{++}$ to the gauge covariant one
$\nabla^{++}={\cal D}^{++}+V^{++}$:
\bea
S_q&=&\int d\zeta^{(-4)} \bar q^+\nabla^{++} q^+\,,
\label{Sq}\\
S_\omega&=&\int d\zeta^{(-4)} \nabla^{++}\omega \nabla^{++}\omega\,.
\label{Somega}
\eea

For the time being we do not specify the representation of gauge group on the
matter fields. We only notice that there is a difference between the $q$ and
$\omega$ hypermultiplet models in this aspect: since the $\omega$ hypermultiplet
is described by a real superfield, it can be naturally placed into a real representation, e.g.
the adjoint representation, while the $q$ hypermultiplet is well suited
for putting it into a complex representation of the gauge group, e.g. the fundamental one.
Actually, there is a duality-sort transformation between two types of the hypermultiplet
\cite{GIOS2}, so this
difference between them is, to some extent, conventional.

Apart from the minimal gauge interaction, one can consider the
hypermultiplet self-interaction. For the model with a single
$q$ hypermultiplet there exists the unique possibility to construct
a quartic SU(2)$_R$ invariant superfield potential \footnote{To prevent a possible confusion, let us recall
that such $q$ superfield ``potentials'', after passing to the physical component fields, give rise to the sigma-model
terms for the latter rather than to a scalar potential and Yukawa-type fermionic couplings \cite{HyperKaehler}.
However, these component potential terms
can appear as an effect of presence of central charges in the supersymmetry algebra.}
\be
S_4=\lambda\int d\zeta^{(-4)} (q^+\bar q^+)^2\,,\qquad
[\lambda]=-1\,.
\label{q4}
\ee
In Section\ \ref{sc-pot} we will show that the self-interaction
(\ref{q4}) emerges as a leading quantum correction in the model of  massive
charged hypermultiplet.

\section{Background field quantization}
The background field method is a powerful tool for studying
the general structure of the quantum effective actions in gauge theories. The basic advantage of
this method is that it gives an opportunity to evaluate the effective action with preserving
the classical gauge invariance on all steps of quantum computations. The idea
of the background field method consists in splitting the initial fields into the classical and quantum parts
and fixing the gauge symmetry only for the quantum fields in the generating functional for the effective
action. For supersymmetric field models the concrete realizations
of such a splitting is a non-trivial task which requires a special study in every case.
For $\cN{=}1$, $d{=}4$ supergauge theories the
background field method is discussed in \cite{GGRS}, \cite{bookBK}.
For $\cN{=}2$, $d{=}4$ supersymmetric theories formulated in terms of (constrained)
$\cN{=}2$ superfields such a method was worked out in \cite{HST}. For the $\cN{=}2$, $d{=}4$
supergauge theories in harmonic superspace
this method was developed in \cite{BBKO}, \cite{non-ren}. Subsequently, it was successfully applied
for studying quantum aspects of these models.

In this section we formulate the background field method
for the $\cN{=}3$, $d{=}3$ Chern-Simons matter
theory with the following general action
\be
S=S_{CS}+S_{q}\,,
\label{Svq}
\ee
where the Chern-Simons and hypermultiplet actions are given by
eqs. (\ref{CS}), (\ref{Sq}).\footnote {Here we do not consider the
$\cN{=}3$ SYM theory (\ref{SYM}) since we concentrate on the conformally
invariant models.}

\subsection{The background field method for $\cN{=}3$ Chern-Simons theory}
The background field method for the $\cN{=}3$, $d{=}3$ Chern-Simons theory is analogous in some points to the one
for the $\cN{=}2$, $d{=}4$ SYM theory
\cite{BBKO}, because the harmonic superspace classical actions in both theories bear a close resemblance to each other.

The classical action in the $\cN{=}3$ Chern-Simons theory (\ref{CS})
is invariant under the gauge transformations (\ref{gauge-tr}). We
split the gauge superfield $V^{++}$ into the `background' $V^{++}$
and `quantum' $v^{++}$ parts,
\be
V^{++}\longrightarrow V^{++}+\kappa v^{++}\,,
\label{splitting}
\ee
where
\be
\frac1{\kappa^2}=\frac{ik}{4\pi}\,.
\ee
Then, the infinitesimal gauge transformations (\ref{gauge-tr}) can
be realized in two different ways:

(i) Background transformations
\be
\delta V^{++}=-{\cal
D}^{++}\lambda-[V^{++},\lambda]=-\nabla^{++}\lambda\,,
\qquad \delta v^{++}=[\lambda,v^{++}]\,;
\label{bg-tr}
\ee

(ii) Quantum transformations
\be
\delta V^{++}=0\,,\qquad
\delta v^{++}=-\frac 1\kappa\nabla^{++}\lambda -[v^{++},\lambda]\,.
\label{q-tr}
\ee
Here the covariant harmonic derivative $\nabla^{++}$ involves the background superfield $V^{++}$.
Upon the splitting (\ref{splitting}), the Chern-Simons action (\ref{CS}) can be
rewritten as (see \cite{EChAYa} for details of such a derivation in the ${\cal N}{=}2, d{=}4$ case)
\be
S_{CS}[V^{++}+v^{++}]=S_{CS}[V^{++}]
-\frac1\kappa\tr\int d\zeta^{(-4)}\, v^{++} W^{++}(V^{++})
+\Delta S_{CS}[V^{++},v^{++}]\,,
\label{5.43}
\ee
where $W^{++}(V^{++})$ is defined in (\ref{W++}) and
\be
\Delta S_{CS}[V^{++},v^{++}]=\tr \sum_{n=2}^\infty\frac{(-\kappa)^{n-2}}{n}
\int d^9z du_1\ldots du_n
\frac{v^{++}_\tau(z,u_1)\ldots v^{++}_\tau(z,u_n)}{
(u^+_1u^+_2)\ldots (u^+_nu^+_1)}\,.
\label{deltaS}
\ee
We introduced $v^{++}_\tau=e^{-\Omega}v^{++}e^{\Omega}$, with
the bridge superfield $\Omega$ being constructed from the background
gauge superfield $V^{++}$ by the rule (\ref{bridge}). The action
(\ref{deltaS}) implicitly depends  on the background superfield
$V^{++}$ via the bridge superfield $\Omega$ which is a complicated
function of $V^{++}$. Every term in (\ref{5.43}) is manifestly
invariant under the background gauge transformations
(\ref{bg-tr}). The second term in (\ref{5.43}) is responsible for
the Chern-Simons equation of motion for the background gauge
superfield which is none other than $W^{++}(V^{++})=0$. This term is
not essential while constructing the effective action.

Within the background field method, it is necessary to fix the gauge only with respect to the
quantum gauge transformations (\ref{q-tr}). The corresponding gauge-fixing function is
\be
{\cal F}^{(4)}=\nabla^{++}v^{++}\,,
\ee
or, being rewritten in the $\tau$ frame,
\be
{\cal F}^{(4)}_{\tau}={\cal D}^{++}v^{++}_\tau=
e^{-\Omega}(\nabla^{++}v^{++})e^{\Omega}
=e^{-\Omega}{\cal F}^{(4)}e^{\Omega}\,.
\ee
Under the quantum gauge transformations (\ref{q-tr}) this function is transformed as
\be
\delta {\cal F}^{(4)}_\tau=-\frac1\kappa e^{-\Omega}
\{ \nabla^{++}(\nabla^{++}\lambda +\kappa[v^{++},\lambda])\}e^\Omega\,.
\ee
The corresponding Faddeev-Popov determinant
\be
\Delta_{FP}[V^{++},v^{++}]={\rm Det}\, \nabla^{++}(\nabla^{++}+\kappa v^{++})
\ee
can be represented by a path integral $\int {\cal D}b{\cal
D}c\,\exp(iS_{FP})$ with two ghost superfields $b$, $c$ in the adjoint representation
of the gauge group and with the ghost-field action
\be
S_{FP}=\tr\int
d\zeta^{(-4)}b\nabla^{++}(\nabla^{++}c+\kappa[v^{++},c])\,.
\label{SFP}
\ee
Putting all these ingredients together, we obtain the following representation
for the effective action,
\be
e^{i\Gamma_{CS}[V^{++}]}=e^{iS_{CS}[V^{++}]}
\int {\cal D}v^{++}{\cal D}b{\cal D}c\,e^{i\Delta S_{CS}[V^{++},v^{++}]
+iS_{FP}[b,c,V^{++},v^{++}]}\delta[{\cal F}^{(4)}-f^{(4)}]\,,
\label{G1}
\ee
where $f^{(4)}$ is an arbitrary Lie algebra valued analytic
function and $\delta[{\cal F}^{(4)}-f^{(4)}]$ is the proper functional
delta-function which fixes the gauge.

To cast (\ref{G1}) in a more useful form, we average it with
the following weight factor
\be
\Delta[V^{++}]\exp\left\{
\frac{-i}{2\alpha}\tr\int d^9z du_1du_2\,
f^{(4)}_\tau(z,u_1)\frac{(u^-_1u^-_2)}{(u^+_1u^+_2)^3}f^{(4)}_\tau(z,u_2)
\right\},
\label{weight}
\ee
where $\alpha$ is an arbitrary parameter. The
functional $\Delta[V^{++}]$ can be found from the condition
\be
1=\Delta[V^{++}]\int {\cal D}f^{(4)}
\exp\left\{
\frac{-i}{2\alpha}\tr\int d^9z du_1du_2\,
f^{(4)}_\tau(z,u_1)\frac{(u^-_1u^-_2)}{(u^+_1u^+_2)^3}f^{(4)}_\tau(z,u_2)
\right\}.
\ee
Hence,
\bea
\Delta^{-1}[V^{++}]&=&\int {\cal D}f^{(4)}
\exp\left\{
\frac{-i}{2\alpha}\tr\int d\zeta^{(-4)}_1 d\zeta^{(-4)}_2
f^{(4)}(\zeta_1)A(\zeta_1,\zeta_2)f^{(4)}(\zeta_2)
\right\}\nn\\
&=&{\rm Det}^{-1/2}\hat{A}\,,
\eea
where $\hat{A}$ is some analytic Lie algebra valued operator with the
kernel $A(\zeta_1,\zeta_2)$. To compute ${\rm Det}\hat{A}$, we represent it
by a functional integral over the analytic superfields,
\be
{\rm Det}^{-1}\hat{A}=\int {\cal D}\chi^{(4)}{\cal D}\rho^{(4)}
\exp\left\{i\,\tr\int d\zeta^{(-4)}_1 d\zeta^{(-4)}_2
\chi^{(4)}(\zeta_1)A(\zeta_1,\zeta_2)\rho^{(4)}(\zeta_2)\right\}
\ee
and perform the following change of functional variables
\be
\rho^{(4)}=(\nabla^{++})^2\sigma\,,\qquad
{\rm Det}\frac{\delta \rho^{(4)}}{\delta \sigma}={\rm
Det}(\nabla^{++})^2\,.
\ee
Then we obtain
\bea
&&\tr\int d\zeta_1^{(-4)}d\zeta_2^{(-4)}\chi^{(4)}(\zeta_1)A(\zeta_1,\zeta_2)
\rho^{(4)}(\zeta_2)\nn\\
&=&\tr\int d^9zdu_1du_2\,\chi^{(4)}_\tau(z,u_1)\frac{(u^-_1u^-_2)}{(u^+_1u^+_2)^3}
{\cal D}^{++}_{(2)}\sigma_\tau(z,u_2)
=\frac12\int d^9z du\,\chi^{(4)}_\tau({\cal
D}^{--})^2\sigma_\tau\nn\\
&=&-\tr\int d\zeta^{(-4)}\chi^{(4)}\hat\Delta\sigma\,,
\eea
where
\be
\hat\Delta=\frac18(D^{++})^2(\nabla^{--})^2\,.
\label{hat-delta}
\ee
As a result, $\Delta[V^{++}]$ can be formally written as
\be
\Delta[V^{++}]={\rm Det}_{(0,0)}^{-1/2}(\nabla^{++})^2
\,{\rm Det}_{(4,0)}^{1/2}\hat\Delta\,,
\ee
where
\bea
{\rm Det}_{(0,0)}^{-1/2}(\nabla^{++})^2
&=&\int{\cal D}\phi\,e^{-iS_{NK}[\phi,V^{++}]}\,,\nn\\
S_{NK}[\phi,V^{++}]&=&-\frac12\tr\int
d\zeta^{(-4)}\nabla^{++}\phi\nabla^{++}\phi
\eea
and
\be
{\rm Det}_{(4,0)}^{-1}\hat\Delta
=\int {\cal D}\chi^{(4)}{\cal D}\sigma\,e^{i\int d\zeta^{(-4)}
\chi^{(4)}\hat\Delta\sigma}\,.
\ee
The analytic real bosonic superfield $\phi$ plays the role of
Nielsen-Kallosh ghost in this theory. The classical action for
this superfield coincides with the $\omega$-hypermultiplet action
(\ref{Somega}).

Upon averaging (\ref{G1}) with the weight factor (\ref{weight}) we
arrive at the following path-integral representation for the
effective action
\be
e^{i\Gamma_{CS}[V^{++}]}=e^{iS_{CS}[V^{++}]}({\rm
Det}^{1/2}_{(4,0)}\hat\Delta)
\int{\cal D}v^{++}{\cal D}b{\cal D}c{\cal D}\phi
e^{iS_Q[v^{++},b,c,\phi,V^{++}]}\,,
\ee
where
\bea
S_Q[v^{++},b,c,\phi,V^{++}]&=&\Delta S_{CS}[V^{++},v^{++}]
+S_{GF}[V^{++},v^{++}]\nn\\&&+S_{FP}[b,c,V^{++},v^{++}]+S_{NK}[\phi,V^{++}]
\,.
\eea
Here $S_{GF}[v^{++},V^{++}]$ is the gauge-fixing contribution to
the quantum action given by
\bea
S_{GF}[V^{++},v^{++}]&=&-\frac{1}{2\alpha}\tr
\int d^9z du_1du_2 ({\cal D}^{++}v^{++}_\tau(z,u_1))\frac{(u^-_1u^-_2)}{(u^+_1u^+_2)^3}
({\cal D}^{++}v^{++}_\tau(z,u_2))\nn\\
&=&-\frac1{2\alpha}\tr\int d^9z du_1du_2\frac{v^{++}_\tau(z,u_1)
v^{++}_\tau(z,u_2)}{(u^+_1u^+_2)^2}\nn\\&&
-\frac{1}{2\alpha}\tr\int d\zeta^{(-4)}v^{++}\hat\Delta
v^{++}\,.
\eea

Let us consider the sum of quadratic in $v^{++}$ parts  of $\Delta
S_{CS}$ and $S_{GF}$,
\be
-\frac{1}{2}\left(1+\frac1\alpha\right)
\tr\int d^9z du_1du_2\frac{v^{++}_\tau(z,u_1)
v^{++}_\tau(z,u_2)}{(u^+_1u^+_2)^2}
-\frac{1}{2\alpha}\tr\int d\zeta^{(-4)}v^{++}\hat\Delta
v^{++}\,.
\label{5.64}
\ee
The first term in (\ref{5.64}) vanishes at $\alpha=-1$, and we will adopt this
choice in what follows. As a result, we arrive at
the following final representation for the effective action
\be
e^{i\Gamma_{CS}[V^{++}]}=e^{iS_{CS}[V^{++}]}({\rm
Det}^{1/2}_{(4,0)}\hat\Delta)
\int {\cal D}v^{++}{\cal D}b{\cal D}c{\cal D}\phi
e^{i(S_2[v^{++},b,c,\phi,V^{++}]+S_{int}[v^{++},b,c,V^{++}])}\,,
\label{eff-act}
\ee
where
\bea
S_2[v^{++},b,c,\phi,V^{++}]&=&
\frac{1}{2}\tr\int d\zeta^{(-4)}v^{++}\hat\Delta
v^{++}+\tr\int d\zeta^{(-4)}b(\nabla^{++})^2c
\nn\\&&
+\frac12\tr\int d\zeta^{(-4)}\phi(\nabla^{++})^2\phi\,,
\label{SS2}
\\
S_{int}[v^{++},b,c,V^{++}]&=&
\tr \sum_{n=3}^\infty\frac{(-1)^n \kappa^{n-2}}{n}
\int d^9z du_1\ldots du_n
\frac{v^{++}_\tau(z,u_1)\ldots v^{++}_\tau(z,u_n)}{
(u^+_1u^+_2)\ldots (u^+_nu^+_1)}\nn\\
&&
-\kappa\,\tr\int d\zeta^{(-4)}\nabla^{++}b[v^{++},c]\,.
\label{Sint}
\eea
The equations (\ref{eff-act})--(\ref{Sint}) completely determine the
structure of perturbative expansion for the effective action in the pure
$\cN{=}3$ Chern-Simons theory in a manifestly supersymmetric and
gauge invariant form.

\subsection{Adding hypermultiplets}

Now we include into considerations the $q$-hypermultiplet superfield
with the following classical action
\be
S_q=\int d\zeta^{(-4)} \bar q^+(\nabla^{++} +\kappa\, v^{++})
q^+\,,
\label{Sqv}
\ee
where $\nabla^{++}$ is the covariant harmonic derivative with the background gauge superfield $V^{++}$.
Here we do not specify the representation of the gauge group on
the hypermultiplet.
We split the hypermultiplet superfields into the background $q^+$,
and quantum ${\bf q}^+$ parts,
\be
q^+\longrightarrow q^++{\bf q}^+\,,\qquad
\bar q^+\longrightarrow \bar q^++\bar {\bf q}^+\,.
\ee
Upon such a splitting, the classical action (\ref{Sqv}) can be rewritten
as a sum of the following four pieces
\be
S_q=S_q[\bar q^+,q^+,V^{++}]+S_{lin}+S_2+S_{int}\,,
\label{3.41_}
\ee
where $S_q[\bar q^+,q^+,V^{++}]$ is given by (\ref{Sq}) and is
constructed solely from the classical fields, while the term $S_{lin}$ is
linear in the quantum fields,
\be
S_{lin}=\int d\zeta^{(4)}(\bar {\bf q}^+\nabla^{++}q^++\bar q^+\nabla^{++}{\bf q}^+
+\bar q^+ \kappa\, v^{++} q^+)\,.
\ee
This term can be omitted since it does not contribute to the
effective action. The pieces $S_2$ and $S_{int}$ in (\ref{3.41_}) correspond, respectively,
to that part of the action which is quadratic in the quantum
superfields, and to the interaction term:
\bea
S_2&=&\int d\zeta^{(-4)}(\bar {\bf q}^+\nabla^{++}{\bf q}^+
+\bar {\bf q}^+\kappa\, v^{++} q^++ \bar q^+\kappa\, v^{++}{\bf q}^+)\,,\\
S_{int}&=&\kappa\int d\zeta^{(-4)}\bar {\bf q}^+ v^{++}{\bf q}^+\,.
\eea

Now we can generalize the generating functional for the effective
action (\ref{eff-act}) to the Chern-Simons matter theory,
\bea
e^{i\Gamma_{CS}[V^{++},\bar q^+,q^+]}&=&e^{i(S_{CS}[V^{++}]+
S_q[\bar q^+,q^+,V^{++}])}({\rm
Det}^{1/2}_{(4,0)}\hat\Delta)\nn\\&&\times
\int {\cal D}v^{++}{\cal D}b{\cal D}c{\cal D}\phi {\cal D}\bar{\bf q}^+
{\cal D}{\bf q}^+
e^{i(S_2+S_{int})}\,,
\label{eff-act1}
\eea
where
\bea
S_2&=&
\frac{1}{2}\tr\int d\zeta^{(-4)}v^{++}\hat\Delta
v^{++}+\tr\int d\zeta^{(-4)}b(\nabla^{++})^2c
+\frac12\tr\int d\zeta^{(-4)}\phi(\nabla^{++})^2\phi\nn\\&&
+\int d\zeta^{(-4)}(\bar {\bf q}^+\nabla^{++}{\bf q}^+
+\kappa\,\bar {\bf q}^+ v^{++} q^++ \kappa\, \bar q^+ v^{++}{\bf q}^+)\,,
\label{SSS2}
\\
S_{int}&=&
\tr \sum_{n=3}^\infty\frac{(-1)^n \kappa^{n-2}}{n}
\int d^9z du_1\ldots du_n
\frac{v^{++}_\tau(z,u_1)\ldots v^{++}_\tau(z,u_n)}{
(u^+_1u^+_2)\ldots (u^+_nu^+_1)}\nn\\
&&
+\kappa\int d\zeta^{(-4)}\bar {\bf q}^+ v^{++}{\bf q}^+
-\kappa\,\tr\int d\zeta^{(-4)}\nabla^{++}b[v^{++},c]\,.
\label{SSint}
\eea

The treatment of the $\omega$ hypermultiplet within the
background field method is quite analogous to the above $q$ hypermultiplet consideration.

\subsection{Gauge and hypermultiplet propagators}
It is seen from the action (\ref{SSS2}) that the gauge and
hypermultiplet propagators are defined by the equations
\bea
\langle v^{++}(1) v^{++}(2)\rangle=G^{(2,2)}(1|2):&\quad&
\hat \Delta G^{(2,2)}(1|2)=\delta_A^{(2,2)}(1|2)\,,
\label{fullGv-eq}\\
\langle \bar {\bf q}^+(1) {\bf q}^+(2)\rangle=G^{(1,1)}(1|2):&&
\nabla^{++}G^{(1,1)}(1|2)=\delta_A^{(3,1)}(1|2)\,,
\label{fullGq-eq}\\
\langle \bar {\bf \omega}(1) {\bf\omega}(2)\rangle=G^{(0,0)}(1|2):&&
(\nabla^{++})^2G^{(0,0)}(1|2)=\delta_A^{(4,0)}(1|2)\,,
\label{fullGom-eq}
\eea
where the analytic delta-function is given by
\be
\delta^{(4-q,q)}_A(1|2)=-\frac14D^{++\alpha}_{(1)}D^{++}_{{(1)}\alpha}
\delta^9(z_1-z_2)\delta^{(-q,q)}(u_1,u_2)\,.
\ee
Here $\delta^{(-q,q)}(u_1,u_2)$ is the standard harmonic delta-function \cite{book}.

The solutions of the equations
(\ref{fullGv-eq}), (\ref{fullGq-eq}) and (\ref{fullGom-eq}) are given by the
following expressions
\bea
G^{(2,2)}(1|2)&=&\frac1{\hat\Delta^2}\hat\Delta
\delta_A^{(2,2)}(1|2)\,,
\label{fullGv}\\
G^{(1,1)}(1|2)&=&-\frac 1{12\,\hat\square}(D^{++}_{(1)})^2
(D^{++}_{(2)})^2
{\cal W}^0_{(1)}
\frac{e^{-\Omega(1)}e^{\Omega(2)}\delta^9(z_1-z_2)}{
(u^+_1u^+_2)^3}\,,
\label{fullGq}\\
G^{(0,0)}(1|2)&=&-\frac 1{12\,\hat\square}(D^{++}_{(1)})^2
(D^{++}_{(2)})^2{\cal W}^0_{(1)}
\frac{e^{-\Omega(1)}e^{\Omega(2)}\delta^9(z_1-z_2)(u^-_1u^-_2)}{
(u^+_1u^+_2)^3}\,,
\label{fullGom}
\eea
where
\be
{\cal W}^0=W^0
+\frac14\hat\Delta+
\frac12\nabla^{0\,\alpha}\nabla^0_\alpha
\label{cW}
\ee
and the operators $\hat\Delta$, $\hat\Delta^2$, $\hat\square$
depending on the background gauge superfield $V^{++}$ will be
specified below.

The operator $\hat \Delta$ was introduced in (\ref{hat-delta}). It
has the following basic commutation relations with the Grassmann and
harmonic derivatives
\be
[D^{++}_\alpha,\hat\Delta]=0\,,\qquad
[\nabla^{++},\hat \Delta]\Phi^{(q)}_A=(1-q)W^{++}\Phi^{(q)}_A\,.
\ee
When acting on the analytic superfields, the operators $\hat\Delta$ and
$\hat\Delta^2$ can be represented as
\bea
\hat\Delta&=&(\nabla^0)^2-W^0-W^{++}\nabla^{--}\,,\\
\hat\Delta^2&=&\nabla^m\nabla_m
+3W^{++}W^{--}+(W^0)^2-((\nabla^0)^2W^0)
+(D^{++\alpha}W^{--})\nabla^0_\alpha
-2W^0(\nabla^0)^2
\nn\\&&
-2(\nabla^{0\alpha}W^{++})\nabla^0_\alpha\nabla^{--}
-2W^{++}(\nabla^0)^2\nabla^{--}
-2W^{++}\nabla^{0\alpha}\nabla^{--}_\alpha
+W^0W^{++}\nabla^{--}
\nn\\&&
+3W^{++}W^0\nabla^{--}
-((\nabla^0)^2W^{++})\nabla^{--}
+W^{++}W^{++}(\nabla^{--})^2
\,.
\label{Delta2}
\eea
Since the expression (\ref{Delta2}) starts with the square $\nabla^m\nabla_m$,
the operator $1/\hat\Delta^2$ in (\ref{fullGv}) is well defined as a power series expansion
around $\nabla^m\nabla_m$.

The hypermultiplet propagators (\ref{fullGq}) and (\ref{fullGom}) involve
the operator
\be
\hat\square=\frac16(D^{++})^2{\cal W}^0(\nabla^{--})^2\,,
\ee
where ${\cal W}^0$ is defined in (\ref{cW}).
This operator $\hat\square$ reveals the following basic properties
\be
[D^{++}_\alpha,\hat\square]=0\,,\qquad
[\nabla^{++},\hat\square]\Phi_A^{(q)}
=(1-q)\left\{
\left((\nabla^0)^2 W^{++}\right)+[W^0,W^{++}] \right\}\Phi_A^{(q)}\,,
\ee
where $\Phi_A^{(q)}$ is some analytic superfield. With making use of these properties,
the operator $\hat\square$ in application to the analytic superfields is reduced to
\bea
\hat\square&=&\nabla^m\nabla_m
+\frac14\{W^{ij},W_{ij} \}
+\frac12[W^{++},W^{--}]
+(D^{++\alpha}W^{--})\nabla^0_\alpha
+(\nabla^{0\alpha}W^{++})\nabla^{--}_\alpha
\nn\\&&
-((\nabla^0)^2W^{++})\nabla^{--}
+[W^{++},W^0]\nabla^{--}
-((\nabla^0)^2W^0)
\,.
\label{cowboy}
\eea
The expression (\ref{cowboy}) starts with $\nabla^m\nabla_m$,
hence the operator $\hat\square^{-1}$ is well defined as a power series expansion.

For the vanishing background field the propagators
(\ref{fullGv}), (\ref{fullGq}) and (\ref{fullGom})
take very simple form,
\bea
G^{(2,2)}_0(1|2)&=&\frac1{\square}(D^0)^2\delta_A^{(2,2)}(1|2)\,,
\label{freeGv}\\
G^{(1,1)}_0(1|2)&=&-\frac1{16\,\square}
(D^{0}_{(1)})^2 (D^{++}_{(1)})^2 (D^{++}_{(2)})^2
\frac{\delta^9(z_1-z_2)}{(u^+_1u^+_2)^3}\,,
\label{freeG}\\
G^{(0,0)}_0(1|2)&=&-\frac1{16\,\square}(D^0_{(1)})^2(D^{++}_{(1)})^2
(D^{++}_{(2)})^2\delta^9(z_1-z_2)\frac{(u^-_1u^-_2)}{(u^+_1u^+_2)^3}\,.
\label{freeOmega}
\eea
Notice that the free $q$- and $\omega$-hypermultiplet propagators are, respectively, antisymmetric and
symmetric with respect to interchanging their arguments,
\be
G_0^{(1,1)}(1|2)=-G_0^{(1,1)}(2|1)\,,\qquad
G_0^{(0,0)}(1|2)=G_0^{(0,0)}(2|1)\,.
\ee
We will use these free propagators in the next Section where some examples
of quantum computations within this approach will be presented.

An alternative representation for the free propagators
(\ref{freeGv}), (\ref{freeG}) and (\ref{freeOmega}) is given by
\bea
G^{(2,2)}_0(1|2)&=&-\frac1{2\pi i}
\frac{1}{\sqrt{2\rho^{\alpha\beta}\rho_{\alpha\beta}}}
[(\theta^{++}_1)^2-2(u^+_1u^-_2)^2
(\theta^{++}_1\theta^{++}_2)\nn\\&&
+(u^+_1u^-_2)^4(\theta^{++}_2)^2]\delta^{(-2,2)}(u_1,u_2)\,,\\
G^{(1,1)}_0(1|2) &=&\frac1{2\pi i}
\frac{(u^+_1u^+_2)}{\sqrt{2\rho^{\alpha\beta}\rho_{\alpha\beta}}}\,,\\
G^{(0,0)}_0(1|2) &=&\frac1{2\pi i}
\frac{(u^+_1u^+_2)(u^-_1u^-_2)}{\sqrt{2\rho^{\alpha\beta}\rho_{\alpha\beta}}}\,,
\eea
where
\bea
\rho^{\alpha\beta}&=&
x_{A\;1}^{\alpha\beta}-x_{A\; 2}^{\alpha\beta}-
2i\theta^{0(\alpha}_1\theta^{0\beta)}_2-
\frac{2i}{(u^+_1u^+_2)}
\bigg[(u^-_1u^-_2)\theta^{++(\alpha}_1\theta^{++\beta)}_2
-(u^-_1u^+_2)\theta^{++(\alpha}_1\theta^{0\beta)}_2
\nn\\&&
-(u^+_1u^-_2)\theta^{0(\alpha}_1\theta^{++\beta)}_2+
(u^-_1u^+_2)\theta^{++(\alpha}_1\theta^{0\beta)}_1+
(u^+_1u^-_2)\theta^{0(\alpha}_2\theta^{++\beta)}_2\bigg]
\eea
is manifestly analytic $\cN{=}3$ supersymmetric interval.

The quantization of the $\cN{=}3$, $d{=}3$ superfield theories was considered
for the first time in the formalism with the ${\rm O}(3)/{\rm O}(2)$ harmonics
in \cite{5}.

\subsection{$\cN{=}3$, $d{=}3$ nonrenormalization theorem}
It is well known that the $\beta$-function for Chern-Simons coupling
in an arbitrary Chern-Simons matter theory is trivial \cite{KP},
the divergences may occur only in the sector of matter fields.
As for supersymmetric Chern-Simons matter theory, one can hope that the
supersymmetry may reduce the degree of such divergences or even ensure
their full cancellation like in the $\cN{=}4$, $d{=}4$ SYM theory. The
nonrenormalization properties of some $\cN{=}1$ and $\cN{=}2$
Chern-Simons matter theories were discussed in \cite{Kazakov}.
It was shown that in the general case such $\cN{=}1$ and $\cN{=}2$ theories
with scale-invariant superpotentials are not free of UV
divergences, but for some particular superpotentials, when the supersymmetry is enhanced to
$\cN{=}6$ or $\cN{=}8$, the cancellation of such divergences may occur \cite{Gus}.

Here we prove the nonrenormalization theorem in general $\cN{=}3$
Chern-Simons matter theory. The general statement is as follows:
The effective action in the $\cN{=}3$ Chern-Simons model (\ref{CS})
with arbitrary number of $q$ and $\omega$ hypermultiplets
(\ref{Sq}), (\ref{Somega}) in an arbitrary representation of gauge group is
completely finite, in the sense that superfield Feynman diagrams contributing to the effective
action show up no any UV quantum divergences.

This statement is very similar to the nonrenormalization theorem
for the $\cN{=}2$, $d{=}4$ supergauge theory \cite{non-ren} which provides the
finiteness of this theory beyond one loop. In fact, this
analogy is even deeper: the form of the Chern-Simons action
(\ref{CS}) is similar to the $\cN{=}2$, $d{=}4$ SYM action (there is a
dimensionless coupling constant in both cases), the
only difference being in the fact that the integration is now performed over the
three-dimensional space-time. The form of classical harmonic superfield Lagrangians
for the $q$ and $\omega$ hypermultiplets is completely the same as in
four dimensions. The details of the background field method for
the $\cN{=}3$ Chern-Simons theory given in the previous section are
analogous to those in the $\cN{=}2, d{=}4$ case \cite{BBKO}. Therefore one
can follow all the steps of proving the four-dimensional
$\cN{=}2$ nonrenormalization theorem in \cite{non-ren} to arrive at the
same conclusion in the
$\cN{=}3$, $d{=}3$ Chern-Simons-matter theory. Of
course, the gauge and matter propagators in $\cN{=}3$,
$d{=}3$ theory are slightly different from their four-dimensional counterparts, and this should be
taken into account in the proof of the nonrenormalization
theorem in the considered case.  In what follows we compute the superficial degree of divergences in
this theory and prove that the UV divergences are absent.

For calculating the superficial degree of divergence we need to know
the structure of superfield propagators for the matter and gauge
superfields. Within the background field method these propagators
are given by the expressions (\ref{fullGq}), (\ref{fullGom}) and (\ref{fullGv}), respectively.
However, for computing the superficial degree of divergence it is
sufficient to know the free propagators
(\ref{freeG}), (\ref{freeOmega}) and (\ref{freeGv}) since all terms which complement these
propagators to the gauge covariant form are only able to diminish the degree
of divergence of a diagram.

Let us consider some background-field dependent supergraph $G$ with
$L$ loops, $P$ propagators and $N_{mat}$ external matter legs. In
the process of computation of the contribution of such a graph $N_D$ covariant spinor
derivatives may hit the external legs as a result of integration
by parts, thereby reducing the degree of divergence of the diagram.
Like in the $\cN{=}2$, $d{=}4$ gauge theory, the superficial degree
of divergence $\omega(G)$ of this graph is given by
\be
\omega(G)=3L-2P+(2P-N_{mat}-3L)-\frac12N_D=-N_{mat}-\frac12N_D\,.
\label{index}
\ee
Here $3L$ is the contribution of loop momenta, $-2P$ comes from
the factors $\square^{-1}$ in the propagators while another $2P$
corresponds to the operators $(D^{++})^2(D^0)^2$ standing in the
numerators of the propagators. We point out that the number $2P$
in the term within round brackets in (\ref{index}) is decreased by
the number $N_{mat}$ because each external matter leg effectively
takes one $(D^{++})^2$ operator to restore full superspace measure
by the rule (\ref{MeaS}). Another negative contribution $-3L$ in this term
appears since for each loop we have to apply the identity
(\ref{delta}) which reduces the number of the covariant spinor
derivatives. Thus we see that any
diagram with external matter legs is automatically finite. For
the diagrams without external matter legs, the last contribution
$-\frac12N_D$ in (\ref{index}) plays the crucial role. This
contribution appears when $N_D$ covariant spinor derivatives hit
the background gauge superfield $V^{++}$. In full analogy with the
$\cN{=}2$, $d{=}4$ supergauge theory, one can argue that $N_D>0$ as a result
of using the background field method. Indeed, within the
background field method the result of computing any
diagram automatically comes out in a gauge covariant form. In other words, it is expressed in
terms of the covariant superfield strengths $W^{ij}$ given in (\ref{main}) and
their covariant spinor derivatives. These derivatives are
expressed in terms of the gauge superfield $V^{++}$ with some number
of covariant spinor derivatives on it (see, e.g., (\ref{V-0}), (\ref{W++})).
This means that these derivatives
should be effectively taken off from the propagators, thereby decreasing the
superficial degree of divergence of the resulting graph by the number $N_D$. As a
result, we arrive at the inequality $\omega(G)<0$, which proves the UV finiteness of all quantum
diagrams in the model under consideration. Some examples of such
one-loop quantum computations will be presented in Section \ref{examples},
just to confirm the proof given here. It is worthwhile to forewarn that all calculations
in Section\ \ref{examples} will be performed with massless propagators for
both the matter and the gauge superfields, which may lead to infrared
divergences like in (\ref{3.32}). However,  such divergences
automatically disappear if one studies the contributions to the
effective action within the background field method, when all the
propagators are effectively massive. This completes our arguments
towards the quantum finiteness of the $\cN{=}3$ Chern-Simons matter
theory.

\subsection{General structure of  the on-shell effective action}

For simplicity, we discuss the general structure of low-energy
on-shell effective action in the Abelian Chern-Simons theory interacting
with $q$ hypermultiplet, \footnote{Here we omit the Chern-Simons
coupling constant for simplicity.}
\be
S=\int d\zeta^{(-4)}(\frac12V^{++}W^{++}+\bar q^+{\cal D}^{++}q^+
+\bar q^+ V^{++}q^+)\,.
\label{3.48}
\ee
The classical equations of motion are given by
\be
\nabla^{++}\bar q^+=0\,,\quad
\nabla^{++}q^+=0\,,\quad W^{++}=\bar q^+ q^+\,.
\ee
These equations mean that in the $\tau$ frame the hypermultiplet
superfields are linear in harmonics,
\be
q^+=u^+_i q^i\,, \qquad
\bar q^+=u^+_i \bar q^i\,,
\ee
while the gauge superfield strength reads
\be
W^{ij}=\bar q^{(i}q^{j)}\,.
\label{3.51_}
\ee

In general, the on-shell effective action can be written as a sum of two
terms expressed as integrals over the analytic subspace and full superspace,
\be
\Gamma=S+\bar\Gamma=S+\int d\zeta^{(-4)}{\cal L}_{\rm analytic}+\int d^9z\,{\cal
L}_{\rm full}\,.
\ee
Here $S$ is the classical action, while $\bar\Gamma$ corresponds
to the quantum corrections.
Since the model (\ref{3.48}) is scale invariant and there is no
room for the conformal anomaly, as soon as there are no any
divergences, the effective action $\bar\Gamma$ should be
scale-invariant as well. However, there exist no any other scale invariant
analytic superspace invariants except for the terms of the
classical action (\ref{3.48}). Therefore the effective action
should receive non-trivial contributions only in the form of integrals over the full
superspace,
\be
\bar\Gamma=\int d^9z\,{\cal L}(q^i, \bar q^i,
W^{ij},\ldots)\,,
\label{3.53}
\ee
where dots stand for the terms with various gauge covariant
derivatives of the superfields $q^i$, $\bar q^i$ and $W^{ij}$
while ${\cal L}$ is some scale-independent gauge invariant function of
its arguments. The gauge invariance of the effective action
(\ref{3.53}) is ensured by the use of the background filed
method.

We should take into account that on shell all the superfield
strengths $W^{ij}$ in (\ref{3.53}) are expressed through the
hypermultiplet superfields by virtue of (\ref{3.51_}). Therefore, on
shell the low-energy effective action can depend on the
hypermultiplet superfields and their derivatives of arbitrary
order,
\be \bar\Gamma=\int d^9z\, {\cal L}(q^i,\bar
q^i,D^{ij}_\alpha q^k,D^{ij}_\alpha \bar q^k,
\partial_{\alpha\beta}q^i,\partial_{\alpha\beta}\bar q^i,\ldots)\,.
\label{3.54}
\ee

In principle, one can look for the pure potential-like terms in the
effective action, i.e., terms containing no derivatives. However,
such terms cannot appear in the effective action (\ref{3.54}).
Indeed, there is the unique ${\rm SU}(2)$ invariant independent superfield
combination $\bar q^i q_i$, but any Lagrangian depending only on $\bar q^i
q_i$ would involve a scale, ${\cal L}={\cal L}(\bar q^i
q_i/\Lambda)$, $[\Lambda]=1$. Therefore, the expansion of the effective action
starts from the terms with derivatives. For instance, the
following terms are admissible in the full superspace Lagrangian,
\be
\frac{D^{ij}_\alpha q^k D_{ij}^\alpha q_k}{(\bar q^i q_i)^2}\,,
\quad
\frac{D^{ij}_\alpha \bar q^k D_{ij}^\alpha \bar q_k}{(\bar q^i q_i)^2}\,,
\quad
\frac{D^{ij}_\alpha q^k D_{ij}^\alpha \bar q_k}{(\bar q^i
q_i)^2}\,.
\ee
Further hints concerning the possible structure of the low-energy
effective action can be gained from the explicit quantum superfield
computations.

\subsection{Effective action in the one-loop approximation}

Let us turn back to the general non-Abelian Chern-Simons theory
interacting with some number of $q$ hypermultiplets with the
action (\ref{Svq}).  Within the background field method the effective action
is given by the generating functional
(\ref{eff-act1}). The one-loop contributions to the effective
action are defined by the quadratic action (\ref{SSS2}),
\be
\Gamma^{(1)}=-\frac i2\Tr_{(4,0)}\ln\hat\Delta-\frac
i2\Tr_{Ad}\ln(\nabla^{++})^2
+\frac i2\Tr\ln\left(
\begin{array}{ccc}
0 & \nabla^{++} & q^+\\
\nabla^{++} & 0 & \bar q^+\\
q^+ & \bar q^+ &\hat\Delta
\end{array}
\right).
\label{3.56_}
\ee
The first term in (\ref{3.56_}) corresponds to $({\rm
Det}_{(4,0)}\hat\Delta)$ in (\ref{eff-act1}) while the second term
\\
$-\frac i2\Tr_{Ad}\ln(\nabla^{++})^2$ is responsible for the
contributions from the ghost superfields which are in the adjoint
representation of the gauge group. The last matrix term in
(\ref{3.56_}) appears from the second line of (\ref{SSS2}) and it
takes into account both the hypermultiplet and gauge superfield
contributions. Making the Cartan-Iwasawa decomposition of this
matrix, we can rewrite the effective action in the following form
\be \Gamma^{(1)}=\frac i2\Tr_{(2,2)}\ln\hat\Delta -\frac
i2\Tr_{(4,0)}\ln\hat\Delta -\frac i2\Tr_{Ad}\ln(\nabla^{++})^2
+\frac i2\Tr\ln H\,, \label{3.57_} \ee where the operator $H$ is
given by \be H=\left(
\begin{array}{cc}
-q^+\frac1{\hat\Delta}q^+ & \nabla^{++}-q^+\frac1{\hat\Delta}\bar
q^+ \\
\nabla^{++}-\bar q^+\frac1{\hat\Delta}q^+ &
- \bar q^+ \frac1{\hat\Delta}\bar q^+
\end{array}
\right).
\ee
The expression (\ref{3.57_}) is the starting point for the
one-loop perturbation theory in the general $\cN{=}3$ Chern-Simons matter
theory.

\section{Examples of supergraph computations}
\label{examples}

\subsection{Hypermultiplet two-point function}
\label{hyp2pt}
Let us consider the $q$-hypermultiplet effective action in the case of
Abelian gauge superfield
\be
\Gamma_{hyp}=i\,\Tr\ln({\cal
D}^{++}+V^{++})=\sum_{n=2}^\infty\Gamma_{hyp,n}\,,\qquad
\Gamma_{hyp,n}=i\frac{(-1)^{n+1}}{n}
\Tr
\left(
\frac1{{\cal D}^{++}}V^{++}
\right)^n.
\label{3.23}
\ee
Explicitly, the two-point function $\Gamma_2$ depicted in Fig.\ 1a,
is given by
\be
\Gamma_{hyp,2}=-\frac i2\int d\zeta^{(-4)}_1 d\zeta^{(-4)}_2
G_0^{(1,1)}(1|2)V^{++}(2)G_0^{(1,1)}(2|1)V^{++}(1)\,.
\label{2pt}
\ee
Next, we apply the expression (\ref{freeG}) for the propagator and
use two $(D^{++})^2$ operators to restore the full $\cN{=}3$ harmonic
superspace measure by the rule (\ref{MeaS}),
\begin{figure}[tb]
\begin{center}
\includegraphics{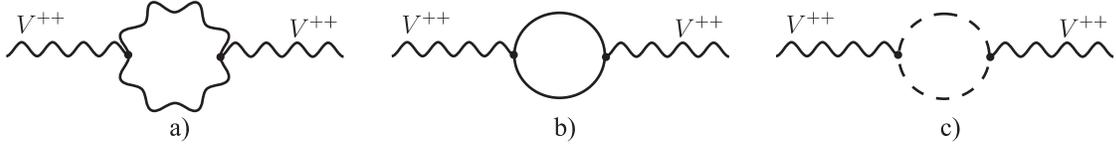}
\caption{Hypermultiplet, gauge superfield and ghost contributions to $\Gamma_2$.}
\end{center}
\end{figure}
\bea
\Gamma_{hyp,2}&=&-\frac{i}{32}\int d^3x_1 d^6\theta_1 d^3x_2
d^6\theta_2 du_1 du_2\, V^{++}(1)V^{++}(2)
\frac1\square (D^0_1)^2\frac{\delta^9(z_1-z_2)}{(u^+_1 u^+_2)^3}\nn\\
&&\times \frac1\square(D^{++}_1)^2 (D^{++}_2)^2 (D^0_2)^2
\frac{\delta^9(z_2-z_1)}{(u^+_2 u^+_1)^3}\,.
\eea
To shrink down the loop over Grassmann variables to a point
we apply the identity
\be
\delta^6(\theta_1-\theta_2)(D^{++}_1)^2 (D^{++}_2)^2 (D^0_2)^2
\delta^9(z_2-z_1)=-16(u^+_1u^+_2)^4
\delta^9(z_1-z_2)\,,
\label{delta}
\ee
and pass to the momentum representation for the superfields,
\be
\Gamma_{hyp,2}=\frac 1{16}\int \frac{d^3p}{(2\pi)^3}  d^6\theta
\frac{ du_1 du_2}{(u^+_1u^+_2)^2}\,\frac1{\sqrt{p^2}}
((D^0_1)^2V^{++}(p,\theta,u_1))V^{++}(-p,\theta, u_2)\,.
\label{3.80}
\ee
In the Abelian case the relation (\ref{V--}) between $V^{++}$ and
$V^{--}$ becomes very simple,
\be
V^{--}(z,u_1)=\int du_2 \frac{V^{++}(z,u_2)}{(u^+_1u^+_2)^2}\,.
\label{V--ab}
\ee
Using this relation, the expression (\ref{3.80}) can be rewritten
as
\bea
\Gamma_{hyp,2}&=&\frac i{16}\int d^3x  d^6\theta du \,
((D^0)^2V^{++}(x,\theta,u))\frac1{\sqrt\square}V^{--}(x,\theta, u)
\nn\\
&=&\frac {i}{16}\int d\zeta^{(-4)} \,
((D^0)^2V^{++})\frac1{\sqrt\square}W^{++}\,.
\label{G2}
\eea
By $1/\sqrt\square$ we denote a non-local operator which acts as
the multiplication by $1/\sqrt{p^2}$ in the momentum
representation. Finally,
it is easy to see that (\ref{G2}) is non other than the Abelian SYM
action with the insertion of the non-local operator
$1/\sqrt\square$,
\be
\Gamma_{hyp,2}=-\frac i{16}\int d\zeta^{(-4)}
W^{++}\frac1{\sqrt\square}W^{++}\,.
\label{G2_}
\ee
A similar result was obtained in the non-supersymmetric
Chern-Simons matter theory \cite{CSW}, as well as in
studying quantum corrections in BLG theory \cite{Gus}.

As a result, the leading contribution to the hypermultiplet effective action
given by the two-point function reproduces the SYM action with the
insertion of non-local operator $1/\sqrt\square$. Of course, such
a non-local operator appears because we do our computations in the
massless theory in which the momentum integral
$\int \frac{d^3k}{k^2(p+k)^2}$ is plagued by the infrared divergence at small
$p$. At zero external momentum, $p=0$, one can regularize this
integral by introducing the parameter $\Lambda$ as a cut-off at small
$k$,
\be
\int \frac{d^3k}{(2\pi)^3}\frac1{k^2(p+k)^2}
\xrightarrow{\mbox{\scriptsize local limit}}
\int \frac{d^3k}{(2\pi)^3}\frac1{k^4}
\rightarrow\frac i{2\pi^2}\int_\Lambda^\infty
\frac{dk}{k^2}=
\frac{i}{2\pi^2}\frac1\Lambda\,.
\label{mom-int1}
\ee
Then, the action (\ref{G2_}) in the local limit is given by
\be
\Gamma_{hyp,2}=\frac1{4\pi^2\Lambda}\int d\zeta^{(-4)}
W^{++}W^{++}\,.
\label{3.32}
\ee
Alternatively, to avoid the regularization of the momentum
integral (\ref{mom-int1}), one can consider the model of massive
hypermultiplet interacting with the Abelian gauge
superfield. Such a model is studied in Section\ \ref{massive-hyper}.

One can easily generalize the result (\ref{G2_}) to the case of a
hypermultiplet in some representation $R$ of non-Abelian gauge group
$G$,
\be
\Gamma_{hyp,2}=-T(R)\frac i{16}\int d\zeta^{(-4)}
W^{++a}_0\frac1{\sqrt\square}W^{++a}_0\,.
\label{G2-ad}
\ee
Here, $\tr(T_R^a T_R^b)=T(R)\delta^{ab}$, $T(\mbox{adjoint})=1$,
and $W_0^{++a}$ is the linear in $V^{++}$ part of
the full non-Abelian superfield strength $W^{++a}$.

\subsection{Gauge and ghost superfield two-point functions}
The next example of quantum computations is represented by the
diagrams b) and c) at Fig.\ 1 which make the leading two-point
contributions to the functional integral (\ref{eff-act}). To study the
pure gauge superfield  diagram a) it is sufficient to consider the
Chern-Simons action (\ref{CS}) up to the cubic term,
\bea
S&=&S_2+S_{gf}+S_3=\frac{ik}{8\pi}\tr\int
d\zeta^{(-4)}V^{++}(D^0)^2
V^{++}\nn\\&&
-\frac{ik}{24\pi}\tr\int d^9z du_1du_2du_3
\frac{V^{++}(z,u_1)[V^{++}(z,u_2),V^{++}(z,u_3)]}{(u^+_1u^+_2)(u^+_2u^+_3)(u^+_3u^+_1)}
\,.
\label{3.86}
\eea
Let us expand the gauge superfields over the generators $T^a$ of
gauge group $G$,\\ $V^{++}=V^{++a}T^a$, so that
\be
[V^{++}(z,u_1),V^{++}(z,u_2)]=V^{++a}(z,u_1)V^{++b}(z,u_2)f^{abc}T^c\,,
\ee
where $f^{abc}$ are the structure constants.
As a result, the action (\ref{3.86}) is rewritten as
\bea
S&=&\frac{ik}{8\pi}\int d\zeta^{(-4)}V^{++a}(D^0)^2
V^{++a}\nn\\&&
-\frac{ik}{24\pi}f^{abc}\int d^9z du_1du_2du_3
\frac{V^{++a}(z,u_1)V^{++b}(z,u_2)V^{++c}(z,u_3)}{(u^+_1u^+_2)(u^+_2u^+_3)(u^+_3u^+_1)}
\,.
\label{3.89}
\eea
The contribution of this action to the one-loop
effective action in the $\cN{=}3$ Chern-Simons theory is as follows
\bea
\Gamma_{CS}&=&\frac i2\Tr\ln\left[
\delta^{ab}\delta_A^{(2,2)}(1|2)-f^{abc}\frac1{16\,\square}
(D^0_{(1)})^2(D^{++}_{(1)})^2(D^{++}_{(2)})^2\delta^9(z_1-z_2)
\right.\nn\\ &&\times \left.
\int du_3\frac
{V^{++c}(z,u_3)}{(u_1^+u^+_2)(u^+_2u^+_3)(u^+_3u^+_1)}
 \right].
 \label{3.90}
\eea
We single out in (\ref{3.90}) the two-point contribution,
\bea
\Gamma_{CS,2}&=&-\frac{i}{16\cdot 64}f^{abc}f^{bad}\int d\zeta_1^{(-4)} d\zeta_2^{(-4)} du_3du_4
\nn\\&&\times
\frac1\square(D^0_{(1)})^2(D^{++}_{(1)})^2(D^{++}_{(2)})^2\delta^9(z_1-z_2)
\frac{V^{++c}(z_1,u_3)}{(u^+_1u^+_2)(u^+_2u^+_3)(u^+_3u^+_1)}
\nn\\&&
\times\frac1\square(D^0_{(2)})^2(D^{++}_{(2)})^2(D^{++}_{(1)})^2\delta^9(z_2-z_1)
\frac{V^{++d}(z_2,u_4)}{
(u^+_2u^+_1)(u^+_1u^+_4)(u^+_4u^+_2)}\,.
\eea
Further computations are analogous to those performed  in the previous
subsection: we restore the full superspace measure by the rule
(\ref{MeaS}) and shrink down the loop over the Grassmann
variables to a point using the identity (\ref{delta}),
\bea
\Gamma_{CS,2}&=& \frac i4f^{abc}f^{abd}\int d^3x_1 d^3x_2 d^6\theta du_1du_2 du_3du_4
\frac1\square\delta^3(x_1-x_2)\frac1\square\delta^3(x_2-x_1)
\nn\\&&
\times\frac{V^{++d}(x_2,\theta,u_4)(D^0_{(1)})^2 V^{++c}(x_1,\theta,u_3)(u^+_1u^+_2)^2}{
(u^+_2u^+_3)(u^+_3u^+_1)(u^+_1u^+_4)(u^+_4u^+_2)}\,.
\eea
To compute the harmonic integrals, we apply the following identity
\be
\int du_1du_2\frac{(u^+_1u^+_2)^2}{(u^+_2u^+_3)(u^+_3u^+_1)(u^+_1u^+_4)(u^+_4u^+_2)}
=-2\frac{(u^-_3u^-_4)}{(u^+_3u^+_4)}\,.
\ee
Passing to the momentum representation and computing the momentum
integral, we find
\be
\Gamma_{CS,2}=\frac 1{16}f^{abc}f^{abd}\int \frac{d^3p}{(2\pi)^3} d^6\theta du_1du_2
\frac1{\sqrt{p^2}}(D^0_{(1)})^2 V^{++c}(p,\theta,u_1)V^{++d}(-p,\theta,u_2)
\frac{(u^-_1u^-_2)}{(u^+_1u^+_2)}\,.
\label{3.94}
\ee
This expression is non-local only in the harmonic variables.

Finally, we consider the ghost field action (\ref{SFP}) which
can be rewritten for the vanishing background field as
\be
S_{gh}=\int d\zeta^{(-4)}[b^aD^{++}D^{++}
c^a+f^{abc}b^aD^{++}V^{++b}c^c]\,.
\ee
The one-loop effective action for the ghost superfields reads
(minus sign is due to the odd statistics of ghost superfields)
\be
\Gamma_{gh}=-i\Tr\ln\left[
\delta^{ab}\delta^{(0,4)}_A(1|2)+f^{abc}V^{++c}(2)D^{++}_{(2)}G^{(0,0)}_0(1|2)
 \right].
\ee
We need only the two-point contribution depicted in
Fig.\ 1c,
\be
\Gamma_{hg,2}=\frac i2\int d\zeta^{(-4)}_{(1)}d\zeta^{(-4)}_{(2)}
f^{abc}f^{bad}V^{++d}(1)V^{++c}(2)D^{++}_{(2)}G^{(0,0)}_0(1|2)
D^{++}_{(1)}G^{(0,0)}_0(2|1)\,.
\ee
Further we assume that the structure constants are normalized in such a way
that $f^{abc}f^{bad}=\delta^{cd}$, i.e., $T(\mbox{adjoint})=1$.
Omitting the details of computations (which are analogous to those
in the previous subsection), we obtain
\be
\Gamma_{hg,2}=\frac 1{16}\int \frac{d^3p}{(2\pi)^3} d^6\theta du_1du_2
\frac1{\sqrt{p^2}}
(D^0_{(1)})^2V^{++a}(p,\theta,u_1)V^{++a}(-p,\theta,u_2)
\frac{(u^-_1u^+_2)(u^+_1u^-_2)}{(u^+_1u^+_2)^2}\,.
\label{3.98}
\ee

Now we sum up the gauge and ghost superfields two-point contributions
(\ref{3.94}), (\ref{3.98}),
\bea
\Gamma_{gauge,2}&=&\Gamma_{CS,2}+\Gamma_{gh,2}\nn\\&=&
-\frac 1{16}\int \frac{d^3p}{(2\pi)^3} d^6\theta du_1du_2
\frac{(D^0_{(1)})^2V^{++a}(p,\theta,u_1)V^{++a}(-p,\theta,u_2)}
{\sqrt{p^2}(u^+_1u^+_2)^2}\,,
\label{3.99}
\eea
where the following identity has been used
\be
\frac{(u^-_1u^+_2)(u^+_1u^-_2)}{(u^+_1u^+_2)^2}-
\frac{(u^-_1u^-_2)}{(u^+_1u^+_2)}=
-\frac1{(u^+_1u^+_2)^2}\,.
\ee
The expression (\ref{3.99}) can be rewritten in the analytic
superspace,
\be
\Gamma_{gauge,2}=\frac i{16}
\int d\zeta^{(-4)}\, W^{++a}_0\frac1{\sqrt\square}W^{++a}_0\,,
\label{3.51}
\ee
where $W^{++a}_0$ is the linear in $V^{++}$ part of
the full non-Abelian superfield strength $W^{++a}$.

Note that the hypermultiplet two-point function (\ref{G2-ad}) has exactly the
same form, but opposite sign. Hence, these two contributions
cancel out each other if one takes $n$ hypermultiplet $q^+_i$ in
representations $R_i$, providing that $\sum_{i=1}^n T(R_i)=1$.
For instance, one $q$-hypermultiplet in the adjoint representation
is sufficient for these two contributions to cancel each other.

A similar cancellation between the hypermultiplet and gauge superfield
two-point functions plays the important role in the $\cN{=}2$, $d{=}4$
gauge theory \cite{GIOS2}, where it is the manifestation of quantum
UV finiteness of the $\cN{=}4$, $d{=}4$ SYM theory. However, in our
case this cancellation is not of the same significance as for the
four-dimensional
models, because all quantum contributions are now divergenceless.
The term (\ref{3.51}) does not contribute to the
Chern-Simons effective action since it vanishes on the
classical equations of motion for the pure gauge superfields.
Moreover, this term is gauge-variant. This was explained in
\cite{Pis} for non-supersymmetric Chern-Simons theory, but this is true
in our case too. Recall that we work in the Fermi-Feynman
gauge, $\alpha{=}-1$, while the authors of \cite{CSW,Gus} used the
Landau gauge $\alpha{=}0$ for which the contributions of the form
(\ref{3.51}) are absent in the pure Chern-Simons theory.

\subsection{Vanishing of tadpoles and hypermultiplet self-energy}
Now we shall consider the tadpole as well as hypermultiplet self-energy
diagrams depicted in Fig.\ 2 and show that their contributions vanish as a consequence of
the properties of Grassmann and harmonic distributions.

The vanishing of the pure gauge superfield diagram a) at Fig.\ 2 is obvious. Indeed, the
gauge superfield propagator (\ref{freeGv}) involves four Grassmann derivatives acting on
the delta function,
\be
G_0^{(2,2)}(1|2)=-\frac1{4\,\square} (D^0)^2(D^{++})^2\delta^9(z_1-z_2)
\delta^{(-2,2)}(u_1,u_2)\,.
\ee
Therefore it vanishes at the coincident points due to the deficit of
Grassmann derivatives. By the same reason vanish similar tadpole
diagrams with more vector legs outgoing from a single point.

The diagrams on Fig.\ 2b), 2c) vanish because of the properties of
harmonics. Indeed, the hypermultiplet propagator (\ref{freeG}) has
six Grassmann derivatives which are necessary to kill all
Grassmann variables of the Grassmann delta function,
\be
(D^0_{(1)})^2(D^{++}_{(1)})^2(D^{++}_{(2)})^2\frac{\delta^6(\theta_1-\theta_2)}{
(u^+_1u^+_2)^3}\Bigg|_{(1)=(2)}
=-16(u^+_1u^+_2)|_{u_1=u_2}=0\,.
\ee
The similar identity can be obtained for the
$\omega$-hypermultiplet propagator (\ref{freeOmega}) which is
responsible for the ghost field contribution depicted in Fig.\ 2b).

\begin{figure}[tb]
\begin{center}
\includegraphics[width=14cm]{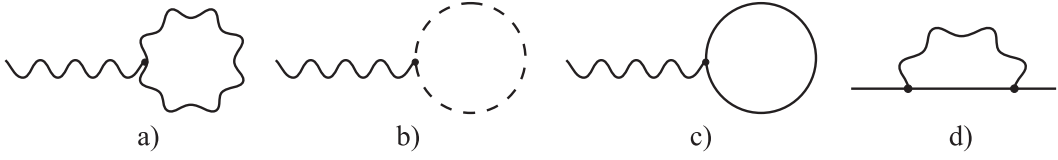}
\caption{Tadpoles and hypermultiplet self-energy diagram.}
\end{center}
\end{figure}
The hypermultiplet self-energy diagram requires a more careful
consideration. Up to a numerical factor, it is given by
\bea
\Gamma_{q\bar q}&\sim&\int d\zeta_1^{(-4)}d\zeta_2^{(-4)}q^+(1)\bar q^+(2)
\frac1\square (D^0_{(1)})^2(D_{(1)}^{++})^2\delta^9(z_1-z_2)
\delta^{(-2,2)}(u_1,u_2)\nn\\&&\times
\frac1\square
(D^0_{(2)})^2(D_{(1)}^{++})^2(D_{(2)}^{++})^2
\frac{\delta^9(z_1-z_2)}{(u^+_1u^+_2)^3}\,.
\eea
Now we restore the full superspace measure and shrink down the
$\theta$-loop using the identity
\be
\delta^6(\theta_1-\theta_2)
(D^0_{(1)})^2(D^0_{(2)})^2(D^{++}_{(1)})^2\delta^9(z_1-z_2)
=-16(u^+_1u^+_2)^2(u^+_1u^-_2)^2\delta^9(z_1-z_2)\,.
\ee
We have exactly six Grassmann derivatives for this identity.
As a result,
\bea
\Gamma_{q\bar q}&\sim&
\int d^3x_1d^3x_2 d^6\theta du_1 du_2\, q^+(x_1,\theta,u_1)
\bar q^+(x_2,\theta,u_2)
\frac{(u^+_1u^-_2)}{(u^+_1u^+_2)}
\delta^{(-2,2)}(u_1,u_2)\nn\\&&
\times
\frac1\square\delta^3(x_1-x_2)\frac1\square\delta^3(x_2-x_1)\,.
\label{4.31_}
\eea
In principle, the harmonic distribution
$\frac{(u^+_1u^-_2)}{(u^+_1u^+_2)}
\delta^{(-2,2)}(u_1,u_2)$ in (\ref{4.31_}) is potentially
dangerous due to the problem of coincident harmonic singularities.
But this problem is resolved here by passing to the analytic
subspace and using the resulting $(D^{++})^2$ operator to produce
extra harmonic factors,
\bea
&&(D^{++}_{(2)})^2 q^+(x,\theta,u_1)
=(u^+_1u^+_2)^2[D^{--}_{(1)}(u^+_1u^+_2)^2
-4D^{--\alpha}_{(1)}D^0_{(1)\alpha}(u^+_1u^+_2)(u^-_1u^+_2)
\nn\\&&\qquad
+4(D^0_{(1)})^2(u^-_1u^-_2)^2]q^+(x,\theta,u_1)\,.
\label{4.32_}
\eea
The factor $(u^+_1u^+_2)^2$ in the r.h.s.  of (\ref{4.32_}) cancels
the denominator of the harmonic distribution in (\ref{4.31_}) and
gives zero due to the identity
$(u^+_1u^+_2)\delta^{(-2,2)}(u_1,u_2)=0$.
As a result, the hypermultiplet self-energy contribution vanishes,
$\Gamma_{q\bar q}=0$.

\section{$\cN{=}3$ supersymmetry with central charges}
\label{massive-hyper}
The $\cN{=}3$ superalgebra without central charges is generated by
the operators (\ref{Q}) with the anticommutation relations
(\ref{Q-algebra}). In this section we shall study an extension
of this superalgebra by the central charge
operators $Z^{ij}$.  The relations (\ref{Q-algebra}) are replaced by the following ones
\be
\{{\mathbb Q}^{ij}_\alpha,{\mathbb Q}^{kl}_\beta\}=
-i(\varepsilon^{ik}\varepsilon^{jl}+\varepsilon^{il}\varepsilon^{jk})
\partial_{\alpha\beta}
-
\frac12\varepsilon_{\alpha\beta}
(\varepsilon^{ik}Z^{jl}+\varepsilon^{jk}Z^{il}
+\varepsilon^{jl}Z^{ik}+\varepsilon^{jl}Z^{ik})\,.
\label{calQ}
\ee
The operators $Z^{ij}$ commute with all other
generators except those of the R-symmetry SU(2) algebra. We will show that just this modified
${\cal N}{=3}$ superalgebra is inherent in the massive hypermultiplet model, in analogy with the four-dimensional
case \cite{IKZ,BK97,Z86}.

\subsection{Massive hypermultiplet model}
Let us consider the Abelian version of the $q$-hypermultiplet model (\ref{Sq})
\be
S_m=\int d\zeta^{(-4)}\bar q^+({\cal D}^{++}+V^{++}_0)q^+\,,
\label{Sm}
\ee
with the background gauge superfield given by
\be
V^{++}_0=3(\theta^{++})^2u^-_iu^-_j Z^{ij}\,,\qquad
Z^{ij}=Z^{ji}=const.
\label{V0}
\ee
One can easily find the relevant bridge superfield $\Omega_0\,$,
\be
\Omega_0=3\theta^{++}\theta^0 u^-_k u^-_lZ^{kl}
+\theta^{--}\theta^0 u^+_ku^+_l Z^{kl}
-\theta^{++}\theta^{--}u^+_ku^-_l Z^{kl}
-2(\theta^0)^2 u^+_ku^-_l Z^{kl}\,,
\ee
as a solution of the equation
\be
{\cal D}^{++}+V^{++}_0=e^{-\Omega_0}{\cal D}^{++}e^{\Omega_0}\,.
\ee
Now, using the relations (\ref{bridge}), we obtain the connections
for the covariant spinor derivatives
\be
{\mathbb D}^{ij}_\alpha=D^{ij}_\alpha+V^{ij}_{0\alpha}\,,\qquad
V^{ij}_{0\alpha}=\frac12\theta^{ik}_\alpha Z^j_k
+\frac12\theta^{jk}_\alpha Z^i_k\,.
\label{4.6}
\ee
These derivatives satisfy the following anticommutation relations
\be
\{{\mathbb D}^{ij}_\alpha,{\mathbb D}^{kl}_\beta \}
=i(\varepsilon^{ik}\varepsilon^{jl}+\varepsilon^{il}\varepsilon^{jk})
\partial_{\alpha\beta}
+
\frac12\varepsilon_{\alpha\beta}
(\varepsilon^{ik}Z^{jl}+\varepsilon^{jk}Z^{il}
+\varepsilon^{jl}Z^{ik}+\varepsilon^{jl}Z^{ik})\,.
\ee
The original supercharges (\ref{Q}) do not anticommute with
(\ref{4.6}). However, one can define the modified supercharges
\be
{\mathbb Q}^{ij}_\alpha=Q^{ij}_\alpha-V^{ij}_{0\alpha}
=Q^{ij}_\alpha-\frac12\theta^{ik}_\alpha Z_k^j
-\frac12\theta^{jk}_\alpha Z_k^i\,,
\label{165}
\ee
so that ${\mathbb Q}^{ij}_\alpha$ anticommute with (\ref{4.6}),
i.e. $\{ {\mathbb Q}^{ij}_\alpha, {\mathbb D}^{kl}_\beta\}=0$. One can easily
check that the operators ${\mathbb Q}^{ij}_\alpha$ satisfy the
anticommutation relations of $\cN{=}3$ superalgebra with central
charges (\ref{calQ}). The central charge operators are realized as
the multiplication by the constants $Z^{ij}$. It is worth noting that these
constant central charges explicitly break the R-symmetry SU(2) down to U(1)$\subset $ SU(2).

It is a non-trivial task to show that the equations of motion in the
model (\ref{Sm}) lead to the mass-shell condition for the
hypermultiplet superfield $q^+$. To this end we introduce the
following notation
\be
Z^{++}=u^+_iu^+_j Z^{ij}\,,\quad
Z^{--}=u^-_iu^-_j Z^{ij}\,,\quad
Z^0=u^+_iu^-_j Z^{ij}
\ee
and
\be
\nnabla^{++}={\cal D}^{++}+V^{++}_0\,,\qquad
\nnabla^{--}={\cal D}^{--}+V^{--}_0\,,
\ee
where
\be
V^{--}_0={\cal D}^{--}\Omega_0=
2\theta^{++}\theta^{--}Z^{--}
+4(\theta^0)^2 Z^{--}
-4\theta^{--}\theta^0 Z^{0}
+(\theta^{--})^2 Z^{++}\,.
\label{V--0}
\ee
The equation of motion in the model (\ref{Sm}) has the following
important corollaries
\be
\nnabla^{++} q^+=0\,,\quad\Rightarrow\quad
(\nnabla^{--})^2 q^+=0\,,\quad\Rightarrow\quad
(D^{++})^2(\nnabla^{--})^2 q^+=0\,.
\label{200}
\ee
Hence, each operator from the set
\be
(D^{++})^2({\mathbb D}^0)^2(\nnabla^{--})^2\,,\quad
(D^{++})^2(\nnabla^{--})^2(D^{++})^2(\nnabla^{--})^2\,,\quad
(D^{++})^2(\nnabla^{--})^2
\ee
annihilates the superfield $q^+$ on-shell. Here
 ${\mathbb D}^0_\alpha=D^0_\alpha+Z^0\theta^0_\alpha-\frac12Z^{--}\theta^{++}_\alpha
-\frac12Z^{++}\theta^{--}_\alpha$. Based on the
important identity for these operators
\bea
&&\frac1{12}(D^{++})^2({\mathbb D}^0)^2(\nnabla^{--})^2
+\frac1{192}(D^{++})^2(\nnabla^{--})^2(D^{++})^2(\nnabla^{--})^2
\nn\\&&\qquad\qquad\qquad\qquad\qquad\qquad\qquad
+\frac16(D^{++})^2Z^0(\nnabla^{--})^2=\square+\frac12Z^{ij}Z_{ij}
\,,
\label{mass-shell}
\eea
which holds in application to the analytic superfields,
we derive the mass-shell condition for the hypermultiplet
superfield,
\be
(\square+m^2)q^+=0\,,\qquad
m^2=\frac12Z^{ij}Z_{ij}\,.
\label{m}
\ee

Thus we have demonstrated that the model (\ref{Sm}) does describe
the massive hypermultiplet model with the mass squared being equal to the
square of the central charge operators. All these considerations
are analogous to those in the four-dimensional $q$-hypermultiplet model.
Minor complications stem from the fact that in the three-dimensional case the central charge $Z^{ij}$ has
${\rm SU}(2)$ indices. It is obvious that such a central charge indeed breaks the
${\rm SU}(2)$ R-symmetry of the $\cN{=}3$ superalgebra down to
${\rm U}(1)$.

The propagator of the massive hypermultiplet
can be easily deduced from the full hypermultiplet propagator
(\ref{fullGq}) by choosing the background superfield strengths to be
constant, $W^{ij}=Z^{ij}$,
\be
G^{(1,1)}_m(1|2)=-\frac1{48}\frac1{\square+m^2}
(D^{++}_{(1)})^2(D^{++}_{(2)})^2
[3({\mathbb D}^0_{(1)})^2+3Z^0-Z^{++}\nnabla^{--}_{(1)}]
\frac{e^{\Omega_0(2)-\Omega_0(1)}\delta^9(z_1-z_2)}{
(u^+_1u^+_2)^3}\,,
\label{Gm1}
\ee
where the mass $m$ is defined in (\ref{m}).

\subsection{$\cN{=}3$ SYM as a quantum correction in the massive hypermultiplet model}
It is well known that the standard Chern-Simons action appears
as a result of computation of the one-loop two-point diagram with
a massive fermion inside and two vector fields on the external
legs \cite{Redlich}. Naively, one could expect that the $\cN{=}3$
supersymmetric version of the Chern-Simons theory (\ref{CS}) can
also be derived from the massive hypermultiplet two-point function
of the form depicted in Fig.\ 1a. Surprisingly, such a computations in the
$\cN{=}3$ supersymmetric theory yields the $\cN{=}3$ SYM action rather than the
Chern-Simons one.

Indeed, consider the model of massive $q$-hypermultiplet interacting
with the background Abelian gauge superfield $V^{++}$,
\be
S_{hyp,m}=\int d\zeta^{(-4)}\bar q^+({\cal D}^{++}+V^{++}_0 +V^{++})q^+\,,
\label{213}
\ee
where $V^{++}_0$ is given by (\ref{V0}). The action (\ref{213})
is invariant with respect to the following Abelian gauge transformations
\be
\delta V^{++}=-{\cal D}^{++}\lambda\,,\quad
\delta q^+ = \lambda q^+\,,\quad
\delta \bar q^+=-\lambda \bar q^+\,,
\ee
$\lambda$ being an analytic superfield gauge parameter.
The formal expression for the massive hypermultiplet two-point
function is given by
\be
\Gamma_2=-\frac i2\int d\zeta^{(-4)}_{(1)}d\zeta^{(-4)}_{(2)}
G_m^{(1,1)}(1|2)V^{++}(2)G_m^{(1,1)}(2|1)V^{++}(1)\,,
\ee
where the massive propagator is defined in (\ref{Gm1}). Subsequent
computations are rather similar to those performed in subection\
\ref{hyp2pt} for the massless hypermultiplet, modulo complications related to the fact
that the expression for the massive hypermultiplet propagator is more
involved as compared to the massless one. As a result, we obtain
\be
\Gamma_2=-\frac i2\frac1{(2\pi)^6}\int d^3p d^6\theta du\,
V^{--}(p,\theta,u)(D^0)^2 V^{++}(-p,\theta,u)
\int d^3k\frac1{k^2-m^2}\frac1{(k+p)^2-m^2}\,.
\ee
Computing the momentum integral at zero momenta $p$, we deduce the
local part of the two-point function in the form
\be
\Gamma_2=\frac1{16\pi m} \int d^9zdu\,
V^{--}(D^0)^2 V^{++}
=-\frac 1{16\pi m} \int d\zeta^{(-4)}
W^{++}W^{++}\,.
\label{4.23}
\ee
As a result, we obtain the $\cN{=}3$ SYM action as a quantum correction
in the massive hypermultiplet model.

The reason why the Chern-Simons term does not appear becomes clear in the
component fields formulation. The hypermultiplet superfield
$q^+$ contains the spinor $\psi^i_\alpha$ which is a doublet of
the ${\rm SU}(2)$ R-symmetry group. That part of the massive hypermultiplet
action (\ref{213}) which involves the spinor field $\psi^i_\alpha$
interacting with the vector field is given by
\be
S_\psi=\frac 12\int d^3x( \bar\psi^{i\alpha}iD_{\alpha\beta}\psi_i^\beta
+\bar\psi^{i\alpha}Z_{ij}\psi^j_\alpha)\,,
\label{Spsi}
\ee
where $D_{\alpha\beta}=\partial_{\alpha\beta}+iA_{\alpha\beta}$.
One can choose the frame with respect to broken ${\rm SU}(2)$ rotations (acting on the doublet indices)
in such a way that the central charge matrix takes the following form
\be
\tilde Z^i_j=\left(
\begin{array}{cc}
im&0\\0&-im
\end{array}
\right).
\ee
The spinors $\psi^1_\alpha$ and $\psi^2_\alpha$ decouple from each other and
the action (\ref{Spsi}) can now be rewritten as a sum of two standard
actions of the massive $3D$ spinors with the opposite masses,
\bea
S_\psi&=&S[\psi^1,m]+S[\psi^2,-m]\,,
\label{4.26}\\
S[\psi,m]&=&\frac i2\int d^3x( \bar\psi^{\alpha}D_{\alpha\beta}\psi^\beta
-m\bar\psi^{\alpha}\psi_{\alpha})\,.
\label{4.27}
\eea
Each of the spinors in (\ref{4.26}) makes the same contribution to the
one-loop two-point function, modulo the sign (see, e.g., \cite{Redlich})
\be
\Gamma_2[A,m]=
\frac 1{8\pi}\frac m{|m|}\int
d^3x\,\varepsilon_{mnp}A^m\partial^n A^p
+\frac 1{48\pi|m|}\int d^3x\,F_{mn}F^{mn}\,,
\ee
where $F_{mn}=\partial_m A_n-\partial_n A_m$.
Therefore the Chern-Simons terms cancel each other in
the full two-point function for the action (\ref{4.26}) and
so the leading contribution is given by the Maxwell term,
\be
\Gamma_2[A,m]+\Gamma_2[A,-m]=
\frac 1{24\pi|m|}\int d^3x\,F_{mn}F^{mn}\,.
\label{4.29}
\ee
The action (\ref{4.23}) is none other than a supersymmetric
generalization of (\ref{4.29}).

The absence of the Chern-Simons term in the hypermultiplet
low-energy effective action can be also understood from simple parity
reasoning. Indeed, the hypermultiplet classical action (\ref{Sq})
is even with respect to the P-reflection while the Chern-Simons
one (\ref{CS}) is odd (see \cite{BILPSZ} for details).
Since there are no any divergences in the
one-loop computation (which, if existing, might produce an anomaly), the resulting
hypermultiplet effective action should be also P-even. Hence, the
Chern-Simons term cannot occur in the hypermultiplet effective
action.

\subsection{Hypermultiplet self-interaction induced by quantum corrections}
\label{sc-pot} It is known that the quartic hypermultiplet
self-interaction (\ref{q4}) appears as a leading quantum
correction in the model of  $\cN{=}2$, $d{=}4$ massive
hypermultiplet interacting with the dynamical Abelian gauge
superfield \cite{IKZ}. In this section we will show that a similar
phenomenon takes place in the $\cN{=}3$, $d{=}3$ gauge theory too.

The classical action of the model under consideration is given by
\be
S_{CS,Ab}+S_{hyp,m}\,,
\ee
where $S_{CS,Ab}$ is the Abelian Chern-Simons action,
\be
S_{CS,Ab}=-\frac{ik}{8\pi}\int d\zeta^{(-4)}V^{++}W^{++}\,,
\ee
while $S_{hyp,m}$ is given by (\ref{213}).
\begin{figure}[tb]
\begin{center}
\includegraphics[height=2.5cm]{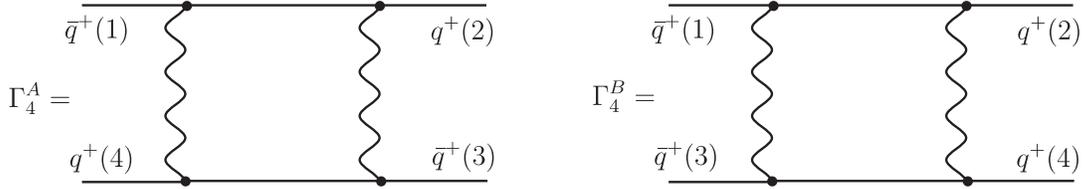}
\caption{Four hypermultiplet contributions.}
\end{center}
\end{figure}
The quartic hypermultiplet self-interaction (\ref{q4}) appears
from the local parts of the diagrams depicted in Fig.\ 3.  Given the
propagators for the gauge superfield (\ref{freeGv}) and for the
massive hypermultiplet (\ref{Gm1}), we represent these
contributions as follows
\bea
\Gamma_4^A&=&-\frac i4\int d\zeta_1^{(-4)}d\zeta_2^{(-4)}d\zeta_3^{(-4)}d\zeta_4^{(-4)}
\bar q^+(1)q^+(2)\bar q^+(3)q^+(4)\nn\\&&
\times G_m^{(1,1)}(1|2)G_0^{(2,2)}(2|3)G_m^{(1,1)}(3|4)G_0^{(2,2)}(4|1)\,,\\
\Gamma_4^B&=&-\frac i4\int d\zeta_1^{(-4)}d\zeta_2^{(-4)}d\zeta_3^{(-4)}d\zeta_4^{(-4)}
\bar q^+(1)q^+(2)\bar q^+(3)q^+(4)\nn\\&&
\times G_m^{(1,1)}(1|2)G_0^{(2,2)}(2|4)G_m^{(1,1)}(3|4)G_0^{(2,2)}(3|1)\,.
\eea
Further computations for $\Gamma_4^A$ and $\Gamma_4^B$ follow the same line.
Therefore we consider in detail only computation of $\Gamma_4^A$.

First, we do the integration over $d\zeta^{(-4)}_3$ and
$d\zeta^{(-4)}_4$ using the analytic delta-function in the gauge superfield
propagator (\ref{freeGv}) and integrate by parts with respect to one of the $(D^0)^2$
operators,
\be
\Gamma_4^A=\frac{4\pi^2 i}{k^2}\int d\zeta_1^{(-4)}d\zeta_2^{(-4)}
\bar  q^+(1) \bar q^+(2)\frac1\square G_m^{(1,1)}(1|2)
[(D^0_{(2)})^2 q^+(2)(D^0_{(1)})^2 q^+(1)\frac1\square
G_m^{(1,1)}(2|1)]\,.
\label{4.35}
\ee
Next, we have to substitute the massive hypermultiplet propagators
(\ref{Gm1}) into this expression. It is important that for deriving the contribution of the form
(\ref{q4}) it is sufficient to take into account only the following term in the massive hypermultiplet
propagator (\ref{Gm1})
\be
G_m^{(1,1)}(1|2)\approx -\frac1{16}\frac1{\square+m^2}
(D^{++}_{(1)})^2(D^{++}_{(2)})^2
(D^0_{(1)})^2\frac{\delta^9(z_1-z_2)}{
(u^+_1u^+_2)^3}\,.
\label{4.36}
\ee
All other terms in the propagator give rise to higher-order contributions involving derivatives.
Substituting (\ref{4.36}) into (\ref{4.35}) and
restoring the full superspace measure, we obtain
\bea
\Gamma_4^A&=&\frac{i\pi^2}{4k^2}\int d^9z_1 d^9z_2 du_1 du_2\,
\bar q^+(2) \frac1{\square(\square+m^2)} \frac{\delta^9(z_1-z_2)}{(u^+_1u^+_2)^3}
\\&&\times
 (D^0_{(1)})^2 \bar q^+(1)
 (D^0_{(2)})^2  q^+(2)
 (D^0_{(1)})^2 q^+(1) (D^0_{(2)})^2 (D^{++}_{(1)})^2(D^{++}_{(2)})^2
  \frac1{\square(\square+m^2)}
\frac{ \delta^9(z_2-z_1)}{(u^+_2u^+_1)^3}\,.\nn
\eea
In this expression, every derivative $D^0_\alpha$ acts on everything
to the right of it. Therefore, there is plenty of terms with
the derivatives $D^0_\alpha$ distributed in different ways among them.
However, a non-trivial result can be generated only by those terms where the operator $(D^0)^2$
is present as a whole. For these terms one can apply the identity (\ref{delta}) to end up with only one $\theta$-integration.
Two other such operators will produce the box operator by the rule
(\ref{D3}). As a result, we are left with the following
expression
\bea
\Gamma_4^A&=&\frac{4\pi^2 i}{k^2}\int d^3x_1 d^3x_2 d^6\theta
\frac{du_1 du_2}{(u^+_1u^+_2)^2}\,
 \frac1{\square(\square+m^2)}\delta^3(x_1-x_2)
 \frac1{\square+m^2}\delta^3(x_2-x_1)
\nn\\&&\times
q^+(2)\bar q^+(2)
 (D^0_{(1)})^2[ \bar q^+(1) q^+(1)]\,.
\eea
Here the term in the second line depends on different $x$'s
and $u$'s, but on the same $\theta$. Next, we pass to the momentum
space and compute the momentum integral in the local limit,
\be
\int \frac{d^3p}{p^2(p^2-m^2)^2}=-\frac {i\pi^2}{m^3}\,,
\ee
thus arriving at
\be
\Gamma_4^A=\frac\pi{2 m^3k^2}\int d^9z
\frac{du_1 du_2}{(u^+_1u^+_2)^2}\,
q^+(z,u_2)\bar q^+(z,u_2)
 (D^0_{(1)})^2[ \bar q^+(z,u_1) q^+(z,u_1)]\,.
\label{4.40}
\ee

The integrand in (\ref{4.40}) contains a harmonic distribution. We
need to single out a local part in this expression in order to get
the contribution of the form (\ref{q4}). For this purpose we follow
the same line as in \cite{IKZ}. We insert the operator ${\cal
D}^0=[\nnabla^{++},\nnabla^{--}]$ under the integral and consider
only the contribution from the term $\nnabla^{++}\nnabla^{--}$ in
this commutator, \footnote{The second term
${\nabla\hspace{-7pt}\nabla}^{--}{\nabla\hspace{-7pt}\nabla}^{++}$
in the commutator contains the operator ${\nabla\hspace{-7pt}\nabla}^{++}$
which hits the hypermultiplet superfields, resulting in the free
massive hypermultiplet equations of motion (\ref{200}).
Therefore, such terms do not contribute to the on-shell
effective action.
Moreover, such terms are non-local with respect to the harmonic
variables while here we are interested in the local contributions to
the effective action.}
\bea \Gamma_4^A&=&\frac\pi{2 m^3k^2}\int d^9z
\frac{du_1 du_2}{(u^+_1u^+_2)^2}\, q^+(z,u_2)\bar q^+(z,u_2) \frac12
{\cal D}^0_{(1)}
 (D^0_{(1)})^2[ \bar q^+(z,u_1) q^+(z,u_1)]\nn\\
 &=&\frac\pi{4 m^3k^2}\int d^9z
\frac{du_1 du_2}{(u^+_1u^+_2)^2}\,
q^+(z,u_2)\bar q^+(z,u_2)
\nn\\&&\qquad\qquad\qquad\qquad\qquad\times
\nnabla^{++}_{(1)}\nnabla^{--}_{(1)}
 (D^0_{(1)})^2[ \bar q^+(z,u_1) q^+(z,u_1)]\,.
\label{4.41}
\eea
We integrate by parts with respect to $\nnabla^{++}$ and use the standard
equation for the harmonic distributions \cite{book},
\be
{\cal D}^{++}_{(1)}\frac1{(u^+_1u^+_2)^2}
={\cal D}^{--}_{(1)}\delta^{(2,-2)}(u_1,u_2)\,,
\ee
which allows us to perform the $u_2$ integration using the harmonic
delta-function,
\bea
\Gamma_4^A&=&\frac\pi{4 m^3k^2}\int d^3x d^6\theta du
\,q^+\bar q^+
(\nnabla^{--})^2
 (D^0)^2[ \bar q^+ q^+]\nn\\
&=&-\frac\pi{16 m^3k^2}\int d\zeta^{(-4)}
q^+\bar q^+
(D^{++})^2 (D^0)^2(\nnabla^{--})^2[ \bar q^+ q^+]\,.
\label{4.42}
\eea
{}From the operator $(\nnabla^{--})^2$ we need only the term
$(V^{--}_0)^2$, where $V^{--}_0$ is given by (\ref{V--0}). Such a
term yields
\be
(D^{++})^2(D^0)^2 (V^{--}_0)^2=32 m^2\,.
\ee
Finally, we find
\be
\Gamma^A_4=-\frac{2\pi}{mk^2}\int d\zeta^{(-4)}
q^+ q^+\bar q^+ \bar q^+ \,.
\ee

One can check that the computation of the second diagram $\Gamma_4^B$ on
Fig.\ 3 yields the same result. Therefore, the final answer for $\Gamma_4$ is as follows
\be
\Gamma_4=-\frac{4\pi}{mk^2}\int d\zeta^{(-4)}
q^+ q^+\bar q^+ \bar q^+ \,.
\ee
It is known that in the $\cN{=}2$, $d{=}4$ hypermultiplet model such
a quartic self-interaction results in a sigma model for the scalar
fields with the target hyper-K\"ahler Taub-NUT metric \cite{HyperKaehler,book}.
The self-interaction (\ref{q4}) gives rise to the same sigma model,
but in the three-dimensional space-time.

\section{Discussion}
In this paper we laid down a basis for the systematic study of the quantum
aspects of three-dimensional $\cN{=}3$ supersymmetric gauge and matter
models in harmonic superspace. We worked out the background field method
for the general $\cN{=}3$ Chern-Simons matter theory. It is a powerful tool
for finding the quantum effective actions directly in $\cN{=}3$, $d{=}3$
harmonic superspace, preserving manifest gauge invariance
and $\cN{=}3$ supersymmetry at each step of the quantum calculations.
The usefulness of this method was illustrated by a simple proof
of the $\cN{=}3$, $d{=}3$ nonrenormalization theorem. Furthermore, we derived the propagators
for the massless and massive hypermultiplets as well as for the Chern-Simons
fields in harmonic superspace and employed them to compute the leading
terms in the quantum two-point and four-point functions.

The derivation of propagators and the calculation of
quantum diagrams in $\cN{=}3$, $d{=}3$ harmonic superspace closely mimic
the analogous considerations in the four-dimensional $\cN{=}2$ harmonic superspace
approach~\cite{GIOS2}. However, in contrast to the four-dimensional case,
there are no one-loop UV divergences in $\cN{=}3$, $d{=}3$ harmonic superspace,
and all diagrams are finite.
Only IR singularities may appear in the massless hypermultiplet
theory, but they can be avoided either by using massive hypermultiplets
or by doing all the calculations within the background field method,
where all propagators are effectively massive.

The massive hypermultiplet model has some new features in
comparison with the four-dimensional theory. As is well known~\cite{FS},
the massive hypermultiplet describes a BPS state, i.e.~it respects
supersymmetry with a central charge equal to the hypermultiplet mass.
The $\cN{=}2$, $d{=}4$ superalgebra has a  central charge (complex or real)
which is a singlet with respect to the R-symmetry group.
Therefore, it breaks the ${\rm U}(2)$ R-symmetry group down
to ${\rm SU}(2)$. In three dimensions, this picture is slightly different.
The $\cN{=}3$, $d{=}3$ superalgebra has a central charge which is a triplet,
breaking the ${\rm SO}(3)\simeq{\rm SU}(2)$ R-symmetry group
down to ${\rm SO}(2)\simeq{\rm U}(1)$.
For this reason, the massive hypermultiplet propagator has a more complicated
form~(\ref{Gm1}) as compared to the four-dimensional case.

A new feature arises when considering quantum contributions in the
massive charged hypermultiplet model. In the $\cN{=}0$ analog of
such a model, i.e.~three-dimensional electrodynamics,
a single massive spinor generates the
Chern-Simons action in the one-loop two-point quantum diagram~\cite{Redlich}.
A similar feature is pertinent to the $\cN{=}1$ and $\cN{=}2$ models~\cite{N2-quant}.
However, the one-loop two-point diagram in the $\cN{=}3$ massive
charged hypermultiplet theory produces the $\cN{=}3$ super Yang-Mills action
rather than the Chern-Simons one as the leading quantum correction.
This may be explained by resorting to a parity argument: the $\cN{=}3$ hypermultiplet is parity-even
while the Chern-Simons term violates parity. Since no anomaly can appear,
the Chern-Simons term is prohibited in the hypermultiplet low-energy
effective action.

Another interesting feature of quantum computations is related to the
one-loop four-point function with four external hypermultiplets in
the model of a massive charged hypermultiplet interacting with a dynamical
Chern-Simons field. We showed that these quantum diagrams produce, as the leading correction,
a quartic hypermultiplet self-interaction which in components
yields the Taub-NUT sigma model for the scalar fields. The same phenomenon
was observed in the four-dimensional case \cite{IKZ}.

Let us outline some further problems which can be studied and
hopefully solved based on the results of the present work. Its
natural continuation is the study of the $\cN{=}3$ superfield
low-energy effective action in the hypermultiplet and the
Chern-Simons theory. So far, there have not been any attempts to
constructing the effective actions in these theories. It is
worthwhile to compare this situation with the $\cN{=}2$, $d{=}4$
supersymmetric models, in which the hypermultiplet and gauge
superfield effective actions have been studied to a large extent
(see, e.g., \cite{EChAYa,d4}). Even more tempting is the
application of our quantum techniques to the $\cN{=}6$ and
$\cN{=}8$ supersymmetric ABJM and BLG models, in order to describe
the quantum-corrected low-energy dynamics of M2 branes in
superstring theory. An important related question concerns the
composite operators for the hypermultiplet superfields in the ABJM
theory. Such operators are relevant for testing the
AdS$_4$/CFT$_3$ version of the general ``gravity/gauge''
correspondence.

\acknowledgments

I.B.S.~is very grateful to I.A.~Bandos and D.P.~Sorokin
for useful discussions and to INFN, Sezione di Padova \&
Dipartimento di Fisica ``Galileo Galilei'', Universita degli Studi
di Padova where this study was started.
The present work is partly supported by RFBR grants, projects
No 08-02-90490 and No 09-02-91349 and by a DFG grant, project No 436 RUS/113/669.
The work of I.L.B, N.G.P and I.B.S is supported also by RFBR
grant, project No 09-02-00078, and
by a grant for LRSS, project No 2553.2008.2.
The work of E.A.I. and B.M.Z. is partly supported by RFBR grants,
projects No 09-02-01209 and No 09-01-93107, and by a grant
of the Heisenberg-Landau program.
I.B.S.\ acknowledges the support from the Dynasty foundation and
from the Alexander von Humboldt Foundation.
N.G.P.\ acknowledges the support from RFBR grant, project No 08-02-00334.

\appendix
\section{Appendix. $\cN{=}3$ harmonic superspace conventions}
\renewcommand{\theequation}{A.\arabic{equation}}
\noindent{\bf Three-dimensional notation}.
We use the Greek letters $\alpha,\beta,\ldots$ to label the
spinorial indices corresponding to the ${\rm SO}(1,2)\simeq {\rm SL}(2,R)$ Lorentz group.
The corresponding
gamma-matrices can be chosen to be real, in particular,
\be
(\gamma^0)_\alpha^\beta=-i\sigma_2=
\left(\begin{array}{cc}
0 & -1 \\ 1 & 0
\end{array} \right),\quad
(\gamma^1)_\alpha^\beta=\sigma_3=
\left(
\begin{array}{cc}
1 & 0 \\0 & -1
\end{array}
\right),\quad
(\gamma^2)_\alpha^\beta=\sigma_1=\left(
\begin{array}{cc}
0 &1\\1 &0
\end{array}
\right).
\label{gamma}
\ee
They satisfy the Clifford algebra
\be
\{\gamma^m,\gamma^n \}=-2\eta^{mn}\,,\qquad
\eta^{mn}={\rm diag}(1,-1,-1)\,,
\ee
and the following orthogonality and completeness relations
\be
(\gamma^m)_{\alpha\beta}(\gamma^n)^{\alpha\beta}=2\eta^{mn}\,,\qquad
(\gamma^m)_{\alpha\beta}(\gamma_m)^{\rho\sigma}=(\delta_\alpha^\rho\delta_\beta^\sigma
+\delta_\alpha^\sigma\delta_\beta^\rho)\,.
\label{A3}
\ee
We raise and lower the spinor indices with the
$\varepsilon$-tensor, e.g., $(\gamma_m)_{\alpha\beta}=\varepsilon_{\alpha\sigma}
(\gamma_m)^\sigma_\beta$, $\varepsilon_{12}=1$.
The products of two and tree gamma-matrices are given by
\bea
(\gamma^m)_\alpha^\rho
(\gamma^n)_\rho^\beta&=&-\eta^{mn}\delta_\alpha^\beta
-\varepsilon^{mnp}(\gamma_p)_\alpha^\beta\,,\\
(\gamma^m)_\alpha^\rho
(\gamma^n)_\rho^\sigma(\gamma^p)_\sigma^\beta&=&
-\eta^{mn}(\gamma^p)_\alpha^\beta
 +\eta^{mp}(\gamma^n)_\alpha^\beta
 -\eta^{np}(\gamma^m)_\alpha^\beta
 +\varepsilon^{mnp}\delta_\alpha^\beta\,,
\eea
where $\varepsilon_{012}=\varepsilon^{012}=1$.

The relations (\ref{A3}) are used to convert any vector index into
a symmetric pair of space-time ones, e.g.,
\bea
&&x^{\alpha\beta}=(\gamma_m)^{\alpha\beta} x^m\,,\qquad
x^m=\frac12(\gamma^m)_{\alpha\beta}x^{\alpha\beta}\,,\nn\\
&&\partial_{\alpha\beta}=(\gamma^m)_{\alpha\beta}\partial_m\,,\qquad
\partial_m=\frac12(\gamma_m)^{\alpha\beta}\partial_{\alpha\beta}\,,
\eea
so that
\be
\partial_m x^n=\delta_m^n\,,\qquad
\partial_{\alpha\beta} x^{\rho\sigma}
=
\delta_\alpha^\rho\delta_\beta^\sigma+\delta_\alpha^\sigma\delta_\beta^\rho
=2\delta_\alpha^{(\rho}  \delta_\beta^{\sigma)}\,.
\ee
\vspace{0.3cm}

\noindent{\bf Superspace and harmonic conventions}. The R-symmetry of $\cN{=}3$
superspace is ${\rm SO}(3)_R\simeq {\rm SU}(2)_R$.
Therefore we label the three
copies of Grassmann variables by a pair of symmetric ${\rm SU}(2)$
indices $i,j$, i.e.,
$\theta_\alpha^{ij}=\theta_\alpha^{ji}$. Thus the $\cN{=}3$
superspace is parametrized by
the following real coordinates in the central basis
\be z=(x^m,
\theta_\alpha^{ij})\,,\quad \overline{x^m}=x^m\,,\quad
 \overline{\theta_\alpha^{ij}}=\theta_{ij\alpha}\,.
\ee
The partial spinor derivatives are defined as follows
\be
\frac\partial{\partial\theta^{ij}_\alpha}\theta^{kl}_\beta
=\delta_\alpha^\beta\,\delta^{k}_{(i}\delta^l_{j)}\,.
\label{A10}
\ee
The covariant spinor derivatives and supercharges read
\be
D^{kj}_\alpha=\frac\partial{\partial\theta^\alpha_{kj}}
 +i\theta^{kj\,\beta}\partial_{\alpha\beta}\,,\qquad
 Q^{kj}_\alpha=
 \frac\partial{\partial\theta^\alpha_{kj}}
 -i\theta^{kj\,\beta}\partial_{\alpha\beta}\,.
\label{Q}
\ee
They satisfy the following  anticommutation relations
\bea
\{D^{ij}_\alpha,D^{kl}_\beta \}&=&i
(\varepsilon^{ik}\varepsilon^{jl}+\varepsilon^{il}\varepsilon^{jk})
\partial_{\alpha\beta}\,,\\
\{Q^{ij}_\alpha,Q^{kl}_\beta \}&=&-i
(\varepsilon^{ik}\varepsilon^{jl}+\varepsilon^{il}\varepsilon^{jk})
\partial_{\alpha\beta}\,.
\label{Q-algebra}
\eea

We use the standard harmonic variables $u^\pm_i$
parametrizing the coset ${\rm SU}(2)_R/{\rm U}(1)_R$ \cite{book}. In particular,
the partial harmonic derivatives  are
\be
\partial^{++}=u^+_i\frac\partial{\partial u^-_i}\,,\quad
\partial^{--}=u^-_i\frac\partial{\partial u^+_i}\,,\quad
\partial^0=[\partial^{++},\partial^{--}]=u^+_i\frac\partial{\partial u^+_i}
-u^-_i\frac\partial{\partial u^-_i}\,.
\label{A12}
\ee
The harmonic projections of the Grassmann $\cN{=}3$ coordinates and spinor
derivatives are defined as follows
\bea
\theta^{ij}_\alpha&\longrightarrow&
(\theta^{++}_\alpha,\theta^{--}_\alpha,\theta^0_\alpha)=
(u^+_iu^+_j\theta^{ij}_\alpha,u^-_iu^-_j\theta^{ij}_\alpha,
u^+_iu^-_j\theta^{ij}_\alpha)\,,\nn\\
D^{ij}_\alpha&\longrightarrow&
(D^{++}_\alpha,D^{--}_\alpha,D^0_\alpha)=
(u^+_iu^+_jD^{ij}_\alpha,u^-_iu^-_jD^{ij}_\alpha,
u^+_iu^-_jD^{ij}_\alpha)\,.
\eea

 The analytic subspace in the full $\cN{=}3$ superspace
is parametrized by the
following coordinates:
\be
\zeta_A=(x^{\alpha\beta}_A,
\theta^{++}_\alpha, \theta^{0}_\alpha, u^\pm_i)\,,
\ee
where
\be
x^{\alpha\beta}_A=(\gamma_m)^{\alpha\beta}x^m_A=x^{\alpha\beta}
+i(\theta^{\alpha++}\theta^{\beta--}+\theta^{\beta++}\theta^{\alpha--})\,.
\ee
The harmonic and Grassmann derivatives in the analytic coordinates are:
\bea
{\cal
D}^{++}&=&\partial^{++}+2i\theta^{++ \alpha}\theta^{0 \beta}
 \partial^A_{\alpha\beta}
 +\theta^{\alpha++}\frac\partial{\partial\theta^{0 \alpha}}
 +2\theta^{0 \alpha}\frac\partial{\partial\theta^{\alpha --}}\,,\nn\\
{\cal D}^{--}&=&\partial^{--}
 -2i\theta^{\alpha --}\theta^{0 \beta}\partial^A_{\alpha\beta}
 +\theta^{\alpha--}\frac\partial{\partial\theta^{0 \alpha}}
 +2\theta^{0 \alpha}\frac\partial{\partial\theta^{++ \alpha}}\,,
\nn\\
{\cal D}^0&=&\partial^0+2\theta^{++ \alpha}\frac\partial{\partial\theta^{++ \alpha}}
-2\theta^{\alpha --}\frac\partial{\partial\theta^{\alpha --}}\,, \quad
[{\cal D}^{++}, {\cal D}^{--}]={\cal D}^0\,,
\label{A16}
\eea
\be
D^{++}_\alpha=\frac{\partial}{\partial
\theta^{\alpha --}}\,,\quad
D^{--}_\alpha=\frac\partial{\partial\theta^{++ \alpha}}
 +2i\theta^{\beta --}\partial^A_{\alpha\beta}\,, \quad
D^0_\alpha=-\frac12\frac\partial{\partial\theta^{0 \alpha}}
+i\theta^{0 \beta}\partial^A_{\alpha\beta}\,,
\label{A17}
\ee
where
$\partial^A_{\alpha\beta}=(\gamma^m)_{\alpha\beta}\frac\partial{\partial
x^m_A}$. They satisfy the following
 relations:
\be
\{D^{++}_\alpha,
D^{--}_\beta\}=2i\partial^A_{\alpha\beta}\,, \quad \{D^{0}_\alpha,
D^{0}_\beta\}=-i\partial^A_{\alpha\beta}\,,
\ee
\be
[{\cal
D}^{\mp\mp}, D^{\pm\pm}_\alpha]=2D^0_\alpha\,, \quad [{\cal D}^{0},
D^{\pm\pm}_\alpha]=\pm 2D^{\pm\pm}_\alpha\,, \quad [{\cal
D}^{\pm\pm}, D^0_\alpha]=D^{\pm\pm}_\alpha\,.
\label{D-alg}
\ee
Some useful identities involving these derivatives are as follows,
\be
(D^0)^2D^0_\alpha=-iD^{0\beta}\partial_{\alpha\beta}\,,\quad
D^0_\alpha(D^0)^2=iD^{0\beta}\partial_{\alpha\beta}\,,\quad
(D^0)^4=\square\,.
\label{D3}
\ee

The integration measures over the full and analytic harmonic superspaces are defined by
\bea
&& d^9z =-\frac1{16}d^3x
(D^{++})^2 (D^{--})^2(D^{0})^2\,,\\
&& d\zeta^{(-4)}=\frac14 d^3x_Adu (D^{--})^2(D^{0})^2\,, \quad d^9z du = -\frac{1}{4} d\zeta^{(-4)}(D^{++})^2\,,
\label{MeaS}
\eea
where $(D^{++})^2=D^{++\alpha}D^{++}_\alpha\,$, etc.
With such conventions, the superspace integration rules are most
simple:
\be
\int d^3x \, f(x)=\int
d^9z(\theta^{++})^2(\theta^{--})^2(\theta^0)^2 f(x)
=\int d\zeta^{(-4)}(\theta^{++})^2(\theta^0)^2 f(x_A)
\ee
for some field $f(x)$.

We denote the special conjugation in the $\cN{=}3$ harmonic superspace by
$\widetilde{\phantom{a}}$,
\bea
\widetilde{(u^\pm_i)}=u^{\pm i}\,,\quad \widetilde{(x^m_A)}=x^m_A\,,\quad \widetilde{(\theta^{\pm\pm}_\alpha)}=
\theta^{\pm\pm}_\alpha\,,\quad \widetilde{(\theta^0_\alpha)}=
\theta^0_\alpha\,.
\eea
It squares to $-1$ on the harmonics and to $1$ on other superspace coordinates.
All bilinear combinations of the Grassmann coordinates are imaginary
\bea
&&\widetilde{[(\theta^{++}_\alpha\theta^0_\beta)]} =-\theta^{++}_\alpha\theta^0_\beta\,,\quad
\widetilde{[(\theta^{++})^2]}=-(\theta^{++})^2\,,\quad
\widetilde{[(\theta^0)^2]}=-(\theta^0)^2\,.
\eea

The conjugation rules for the spinor and harmonic derivatives are
\bea
\widetilde{(D^0_\alpha\Phi)}=-D^0_\alpha\tilde\Phi\,,\quad
\widetilde{[(D^0)^2\Phi]}=-(D^0)^2\tilde\Phi\,,\quad \widetilde{(\cD^{++}\Phi)}=
\cD^{++}\tilde\Phi\,,
\eea
where $\Phi$ and $\tilde\Phi$ are even superfields.

The analytic superspace measure is real,
$\widetilde{d\zeta^{(-4)}}=d\zeta^{(-4)}$, while the full
superspace measure is imaginary,
$\widetilde{d^9z}=-d^9z$.

\end{document}